\documentclass[10pt,preprint]{aastex}
\shorttitle{NANTEN GMC survey in the LMC}
\shortauthors{Fukui et al.}
\begin{document}
\title{The Second Survey of the Molecular Clouds in the Large Magellanic Cloud by NANTEN I:
Catalog of Molecular Clouds}
\author{Fukui, Y.\altaffilmark{1},
Kawamura, A.\altaffilmark{1},
Minamidani, T.\altaffilmark{1},
Mizuno, Y.\altaffilmark{1},
Kanai, Y.\altaffilmark{1},
Mizuno, N.\altaffilmark{1},\\
Onishi, T.\altaffilmark{1},
Yonekura, Y.\altaffilmark{2},
Mizuno, A.\altaffilmark{3},
Ogawa, H.\altaffilmark{2}, \&
Rubio, M.\altaffilmark{4}
}

\altaffiltext{1}{Department of Astrophysics, Nagoya University, Furocho, Chikusaku, Nagoya 464-8602, Japan}
\altaffiltext{2}{Department of Physical Science, Graduate School of Science, 
Osaka Prefecture University,
1-1 Gakuen-cho, Nakaku, Sakai, Osaka 599-8531, Japan}
\altaffiltext{3}{Solar-terrestrial Environment Laboratory, Nagoya University, 
Furocho, Chikusaku, Nagoya 464-8601, Japan}
\altaffiltext{4}{Departamento de Astronomia, Universidad de Chile, Casilla 36-D, Santiago, Chile}
\email{fukui@a.phys.nagoya-u.ac.jp, kawamura@a.phys.nagoya-u.ac.jp}

\begin{abstract}
The second survey of the molecular clouds in $^{12}$CO ($J$ = 1--0) 
was carried out in the Large Magellanic Cloud by NANTEN. 
The sensitivity of this survey is twice as high as
that of the previous NANTEN survey, leading to a detection of
molecular clouds with $M_{\rm CO} \ga 2 \times 10^{4} M_{\sun}$.
We identified 272 molecular clouds, 230 of which are detected at
three or more observed positions.
We derived the physical properties, such as size, line width, virial mass, 
of the 164 GMCs which have an extent more than the beam size of NANTEN
in both the major and minor axes. 
The CO luminosity and virial mass of the clouds show a good correlation of
$M_{\rm VIR} \propto L_{\rm CO}^{1.1 \pm 0.1}$ 
with a Spearman rank correlation of 0.8
suggesting that the clouds are in nearly virial equilibrium.
Assuming the clouds are in virial equilibrium, we derived an 
$X_{\rm CO}$-factor to be $\sim 7 \times 10^{20}$  cm$^{-2}$
(K km s$^{-1}$)$^{-1}$.
The mass spectrum of the clouds is fitted well by a power law of
$N_{\rm cloud}$($>M_{\rm CO}$) $\propto M_{\rm CO}^{-0.75 \pm 0.06}$ above the
completeness limit of $5 \times 10^{4} M_{\sun}$. 
The slope of the mass spectrum becomes steeper if we fit only the massive
clouds; e.g., $N_{\rm cloud}$($>M_{\rm CO}$) $\propto M_{\rm CO}^{-1.2 \pm 0.2}$ for $M_{\rm CO} \ge 3 \times 10^{5} M_{\sun}$.
\end{abstract}
\keywords{Magellanic Clouds --- ISM: structure --- ISM: molecules --- 
galaxies: ISM --- radio lines: ISM --- stars: formation}

\section{Introduction}
Star formation requires cool and high-density ISM in which most of the hydrogen is in molecular form so that the studies of distribution and properties of the molecular clouds are very important to understand the star formation processes. Emission from the tracer CO molecule has been widely used to estimate the distribution and amount of H$_2$ in the Galaxy and other galaxies. Nevertheless, the CO lines in dwarf irregular galaxies are relatively weak (e.g., Ohta et al.\ 1993), making the surveys of molecular clouds toward the dwarf galaxies difficult. 

The Large Magellanic Cloud (LMC) is one of the nearest galaxies to our own. Studies of this galaxy have provided valuable information for our understanding of the universe and galaxies concerning various aspects, including the evolution of stars and stellar clusters, owing to its unrivaled closeness to the solar system ($D \sim$  50 kpc) and favorable viewing angle.  
The LMC also provides a unique opportunity for studying molecular clouds and star formation in galaxies whose environment is different from that in the Galaxy.  In the LMC, the gas-to-dust ratio is $\sim$ 4 times higher (Koornneef 1982) and the metal abundance is about $\sim$ 3 to 4 times lower (Rolleston, Trundle, \& Dufton 2002; Dufour 1984) than that of the Galaxy. 

For these reasons, the LMC has been surveyed at a wide variety of wavelengths. H\,{\scriptsize I} maps of moderate ($\sim$15$'$) to high ($\sim$1$'$) angular resolution have been obtained with the Parkes 64 m telescope (McGee \& Milton 1966; Rohlfs et al.\ 1984; Luks \& Rohlfs 1992) and with the Australian Telescope Compact Array (e.g., Kim et al.\ 1998), respectively. 
These studies have shown that the H\,{\scriptsize I} distribution is dominated by many features, like filaments, shells, or holes. The ionized-gas content has been investigated with the aid of H$\alpha$ photographs (e.g., Henize 1956; Davies et al.\ 1976; Meaburn 1980; Kennicutt \& Hodge 1986)
and the radio continuum (e.g., Haynes et al.\ 1991; Filipovic et al.\ 1996; Dickel et al.\ 2005). The H$\alpha$ and radio continuum images show a variety of interstellar shells, ranging from small supernova remnants to large supergiant shells more than a few 100 pc across. 
The stellar contents of the LMC have been widely studied by photometry of the stars
from NIR to optical bands (e.g., Ita et al.\ 2004, Zaritsky et al.\ 2004, Blum et al.\ 2006,
Kato et al.\ 2006).
Especially, the stellar clusters and associations in the LMC have been surveyed and cataloged by many authors (e.g., Lucke \& Hodge 1970; Hodge 1988). Bica et al.\ (1996) estimated the ages of 504 stellar clusters and 120 associations based on their color indices in the $UBV$ bands. 
Soft X-ray images of the LMC have also been obtained by the ROSAT satellite (Snowden \& Petre 1994), revealing a variety of discrete sources, like supernova remnants, X-ray binaries, supersoft sources, and also diffuse sources associated with superbubbles and supergiant shells.
Recent Spitzer observations cover $\sim 7\arcdeg \times 7\arcdeg$ of the entire LMC
revealing the distribution and properties of the dust, YSOs, and evolved stars (Mexiner et al.\ 2006).

Studies of molecular gas in the LMC began with the observations with either low angular resolution or a small spatial coverage. Cohen et al.\ (1988) obtained the first complete CO map of the LMC with the southern CfA 1.2 m telescope at CTIO. However, the survey was limited by the low spatial resolution, $8.\arcmin8 $ corresponding to 130 pc at the distance of the LMC.
High-resolution CO observations of selected regions by the SEST 15 m telescope have been performed in the LMC (e.g., Israel et al.\ 1986; Johansson et al.\ 1994; Caldwell \& Kutner 1996; Kutner et al.\ 1997; Johansson et al.\ 1998; Israel et al.\ 2003).  They have mapped some of the well-known HII regions, for example, 30 Doradus, N 11, and the molecular cloud complex extending some 2 kpc south of 30 Doradus. Although these observations revealed detailed structure of the individual molecular clouds at a linear resolution of less than 10 pc, they are limited in spatial coverage, 
about one square degree.  

Fukui et al.\ (1999) carried out and completed a survey of the LMC in CO $J =$ 1--0 by NANTEN, a 4 m radio telescope installed at the Las Campanas Observatory, Chile to reveal a molecular gas distribution with resolution high enough to identify individual molecular clouds in the LMC. 
The 3 $\sigma$ noise level of the velocity-integrated intensity was $\sim$ 1.8 K km s$^{-1}$. This corresponds to $N({\rm H}_2) \sim 1.3 \times 10^{21}$ cm$^{-2}$, by using a conversion factor of $X_{\rm CO} = 7 \times 10^{20}$ cm$^{-2}$(K km s$^{-1}$)$^{-1}$ (see section 4.2).
The first results are presented in Fukui et al.\ (1999) with particular emphasis on the formation of populous clusters. 
The catalog of 107 molecular clouds and a comparison of these molecular clouds with HI distribution are described by Mizuno et al.\ (2001b).  A comparison with the young stellar clusters and HII regions are the subjects of Yamaguchi et al.\ (2001c). 
The comparisons of the molecular clouds with HI gas and infrared emission by IRAS were made by Sakon et al.\ (2006) and Hibi et al.\ (2006).

In order to have more comprehensive understanding of the distribution and properties of the molecular clouds in the LMC, a survey in CO $J =$ 1--0 with a sensitivity higher than that of the first survey has been carried out since 1999.
Preliminary results focussing on the mass spectrum of this survey are presented by Fukui et al.\ (2001) and on the comparison of the molecular clouds with super giantshells by Yamaguchi et al.\ (2001a and 2001b). Hughes et al.\ (2006) made a comparison of Radio, FIR, HI and CO in the LMC and revealed the correlation of radio and FIR distributions.

In this paper, we present a catalog and properties of the molecular clouds from the complete dataset of the second survey.
Comparisons of the molecular clouds with the indication of the formation of clusters and 
massive stars detected in optical or radio are found in elsewhere (Fukui 2005; Kawamura et al. 2006; Kawamura et al.\ 2008, ``Paper II", hereafter). 
Fukui (2007) introduces a comparison of the molecular clouds with HI gas distribution.
The detailed studies of the HI and molecular gas distribution to seek for the molecular
cloud formation will be presented in Fukui et al. (2008, ``Paper III", hereafter).
In this paper, the survey by NANTEN is described in Section 2 and the spatial distribution and a catalog of molecular clouds are presented in Section 3.  We also discuss the correlations among cloud properties, such as $L_{\rm CO}$ and the virial mass of the molecular clouds and mass spectrum in Section 3.
A comparison of the cloud properties from the first survey and 
the discussion of the CO to $N$(H$_{2}$) conversion factor are 
presented in Section 4. Section 5 summarizes the paper.

\section{Observations}

We carried out sensitive CO($J=$1--0) observations toward the LMC by NANTEN, a 4 m radio telescope of Nagoya University at Las Campanas Observatory, Chile. The half-power beam width was $2\arcmin.6$ at 115 GHz. The telescope had a 4 K cooled Nb superconductor-insulator-superconductor mixer receiver, which provided a system noise temperature of $\sim$ 170--270 K, including the atmosphere toward the zenith. 

The spectrometers were two acoust-optical spectrometers with 2048 channels; one had a velocity coverage and a resolution of 100 km s$^{-1}$ and 0.1 km s$^{-1}$, respectively.  The other had a velocity coverage and a resolution of 650 km s$^{-1}$ and 0.65 km s$^{-1}$, respectively. The pointing accuracy was better than $20''$, as checked by optical observations of stars with a CCD camera attached to the telescope, as well as by radio observations of Jupiter, Venus, and the edge of the Sun. Further details about the telescope and related instruments are given by Ogawa et al.\ (1990) and by Fukui and Sakakibara (1992). The spectral intensities were calibrated by employing the standard room-temperature chopper wheel technique (Kutner \& Ulich 1981). An absolute intensity calibration was made by observing Orion-KL [R.A. (B1950) = $5^{\rm h}32^{\rm m}47^{\rm s}\hspace{-3pt}.\hspace{2pt}0$, Decl.\ (B1950) = $-$5$^{\circ}$24$'$$21''$]  by assuming its absolute temperature $T_{\rm R}^{*}$ to be 65 K. We also observed the strongest peak position of the LMC, N 159 [R.A. (B1950) = $5^{\rm h}40^{\rm m}1^{\rm s}\hspace{-3pt}.\hspace{2pt}5$, Decl.\ (B1950) = $-$69$^{\circ}$47$' $$2''$] every 2 hours to confirm the stability of the system.

The observed region covers $\sim 30$ square degrees where the molecular clouds were detected by the 1st survey. Figure 1 shows the observed region from 1998 April to 2003 August, superposed on the integrate intensity map of the CO from the 1st survey (Fukui et al.\ 1999; Mizuno et al.\ 2001b). In total, about 26,900 positions were observed in Equatorial coordinate (B1950) in position switching. The observed grid spacing was 2$\arcmin$ (corresponding to $\sim$ 30 pc at a distance of the LMC, 50 kpc) with a 2.6$\arcmin$ beam ($\sim$ 40 pc).
Out of the $\sim$ 26,900 positions, 6,229 were observed with the narrow band spectrometer for $\sim100$ days from April to November 1998, while the rest were observed with the wide band spectrometer for $\sim 300$ days after March 1999. In this paper, all the spectra observed by the narrow band spectrometer are smoothed to 0.65 km s$^{-1}$ resolution for the reduction to have a uniform velocity resolution throughout the map. The rms noise fluctuations were $\sim$ 0.07 K at a velocity resolution of 0.65 km s$^{-1}$ with $\sim$ 3 minutes integration for an on-position.   
The typical 3 $\sigma$ noise level of the velocity-integrated intensity was $\sim$ 1.2 K km s$^{-1}$. This corresponds to $N({\rm H}_2) \sim 8 \times 10^{20}$ cm$^{-2}$, by using a conversion factor of $X_{\rm CO} = 7 \times 10^{20}$ cm$^{-2}$(K km s$^{-1}$)$^{-1}$ (see section 4.2).

\section{Results}

\subsection{Overall distribution}

Out of the $\sim 26,900$ observed positions, significant $^{12}$CO ($J$ = 1--0) emission, with an integrated intensity greater than 1.2 K km s$^{-1}$ (the $\sim 3 \sigma$ noise level), 
was detected at about 1,300 positions, which corresponds to $\sim 5 \%$ of the total observed positions. The total velocity-integrated intensity distribution of the molecular gas is shown in 
Figure \ref{fig:ii}. 

The mass of the molecular gas in total is $\sim 5 \times 10^{7} M_{\sun}$
if we use the CO luminosity to hydrogen column density conversion factor, 
$X_{\rm CO}$-factor,
to be $\sim 7 \times 10^{20}$  cm$^{-2}$ (K km s$^{-1}$)$^{-1}$ (see section 4).
The CO distribution of the LMC is found to be clumpy with several large molecular cloud complexes contrary to the HI gas distribution, which is composed of many filamentary and shell-like structure (e.g., Kim et al.\ 1999). 
These clumps of the molecular gas tend to be detected toward the intensity peak of the HI gas as shown in Blitz et al. (2006), Fukui (2007), and Paper III.
The cloud complex south of 30 Doradus at $\alpha$(J2000)  $\sim 5^{\rm h}40^{\rm m}$ and $\delta$(J2000) from $\sim -71^{\circ}$ to $-69^{\circ} 30 \arcmin$ 
is remarkable 
stretching in a nearly straight line from the north to south, as already noted by the previous CO observations (Cohen et al.\ 1988; Fukui et al. 1999; Mizuno et al.\ 2001b).  The current survey shows that this molecular cloud complex, ``the molecular ridge'', is actually connected to one another by a
low-density molecular gas, while the 1st survey traced only the densest regions 
and identified the molecular ridge to consist of 
two or three discrete entities.  
The arc-like distribution of molecular clouds along the southeastern optical edge of the galaxy (``CO Arc'' in Fukui et al.\ [1999] and Fukui [2002]) is also clearly seen. The current sensitive survey confirms that this ``CO Arc'' indeed clearly represent an arc-like edge of the molecular gas distribution in this eastern boundary of the LMC.

Other CO clouds are distributed over the observed area with moderate concentration towards 
several prominent H\,{\scriptsize II} regions (e.g., N 44 at $\alpha$(J2000) $\sim$ $5^{\rm h}22^{\rm m}$ and $\delta$(J2000) $\sim$ $-68^{\circ}$, N11 at $\alpha$(J2000) $\sim$ $4^{\rm h}55^{\rm m}$ and $\delta$(J2000) $\sim$ $-66\arcdeg 30\arcmin$) and the ``bar'' (e.g., clouds around $\alpha$ $\sim$ $5^{\rm h}20^{\rm m}$ and $\delta$ $\sim$ $-70^{\circ}$). The small molecular clouds are more clearly observable in the current survey, especially, toward a supergiant shell, LMC4, at
$\alpha$(J2000) $\sim$ $5^{\rm h}20^{\rm m}$ -- $5^{\rm h}40^{\rm m}$ and 
$\delta$(J2000) $\sim$ $-68\arcdeg$ -- $-66\arcdeg$(see also Yamauguchi et al.\ 2001a).
A detailed comparison of the molecular gas with the indicators of star formation, such as H $\alpha$ or radio continuum will be discussed in elsewhere (Paper II).

Figure \ref{fig:iihist} is a histogram of the integrated intensity, $I.I.$, of the
each observed position with $I.I. > 0.4$ K km s$^{-1}$ ($\sim 1\sigma$ noise level).
About 980 observed positions have $N$(H$_{2}$) $> 10^{21}$ cm$^{-2}$,
while only 4 positions have $N$(H$_{2}$) $> 10^{22}$ cm$^{-2}$.

Figure \ref{fig:sddist}a shows the radial distribution of CO emission;
the surface density, $\Sigma$, is derived by integrating 
the CO luminosity within annuli spaced by 4$\arcmin$ 
and then divided by an area of the annuli.
The center used is 
$\alpha$(J2000)$=5^{h}17.6^{m}$, 
$\delta$(J2000)$=-69\arcdeg2^{'}$
determined from the kinematics of the HI observations
by Kim et al.\ (1998). 
To see the angular distribution of the CO emission,
the distribution of the surface density, $\Sigma$, 
derived by integrating CO luminosity over 
a sector with a $10\arcdeg$ width and then divided by an observed area of the sector
is also shown in Figure \ref{fig:sddist}b.
The CO luminosity to mass conversion is carried out by assuming a 
conversion factor, $X_{\rm CO}$, of $7 \times 10^{20}$ 
cm$^{-2}$(K km s$^{-1}$)$^{-1}$ (see Section 4.2) for both.
Figure \ref{fig:sddist} indicates that the radial profile of the molecular gas
decreases moderately along the galacto-centric distance
as is also seen in the nearby spiral galaxies (e.g., Wong et al.\ 2002),
although the profile does not really fit to a power law 
as they do for the spiral galaxies by Wong et al. (2000).
It is interesting to note the sharp enhancement 
of the surface density around 2 kpc.
This enhancement is due to the molecular cloud complexes,
the molecular ridge, N11, and N44. 
This enhancement is also seen in the angular distribution,
especially at about $120 \arcdeg$ due to the molecular ridge.
Compared with the nearby galaxies by Wong et al.\ (2002),
the local enhancement of the molecular gas is more conspicuous
in the LMC.

Figure \ref{fig:ch1}a--l are a series of channel maps to show the velocity distribution; they present channel maps with velocities at (a) 200--210 km s$^{-1}$, (b) 210--220 km s$^{-1}$, (c) 220--230 km s$^{-1}$, (d) 230--240 km s$^{-1}$, (e) 240--250 km s$^{-1}$, (f) 250--260 km s$^{-1}$, (g) 260--270 km s$^{-1}$, (h) 270--280 km s$^{-1}$, (i) 280--290 km s$^{-1}$, (j) 290--300 km s$^{-1}$, (k) 300--310 km s$^{-1}$,  and (l) 310--320 km s$^{-1}$, respectively. 

In the velocity range 200--240 km s$^{-1}$, the CO Arc in the southeastern boundary of the LMC and the molecular ridge from the south of 30 Doradus are prominent. The two molecular-cloud complexes associated with active star-forming regions N 11 and N 44 are also prominent in the velocity range between 270--290 km s$^{-1}$. A systematic velocity gradient from the southeast to the northwest has been known from the 1st survey (Mizuno et al.\ 2001b). Our sensitive survey shows a clear feature in the northeast at high velocities, corresponding to LMC 4, in addition to the overall velocity gradient already observed in the 1st survey.
Detailed comparisons of the velocity and spatial distributions of the HI and CO emission
will be presented in Paper III.

\subsection{Identification of CO molecular clouds}

To study the properties of the molecular gas in the LMC, individual cloud was identified by the cloud finding algorithm, fitstoprops (Rosolowsky \& Leroy 2006). First, the intensity data cube is converted to a signal-to-noise data cube by dividing through by the noise at each position to search for significant emission by clipping the maps at a constant signal-to-noise ratio;
a constant signal-to-noise ratio is used instead of a constant flux density threshold although the sensitivity variation across the map is not large (Figure \ref{fig:rms}). Then, pairs of adjacent velocity channels with normalized flux greater than 3 were searched for. For each pair, these velocity channels and all contiguous data pixels with normalized flux greater than 2 were assigned to a candidate molecular cloud.
This process was continued until all the pairs of adjacent $2$ channels with normalized flux greater than 3 have been identified as candidate clouds. 

We identified 272 clouds, of which 230 were detected at more than two observed positions.
In this paper, we shall call these 230 clouds with more than two observed positions ``the GMCs".
The position and intensity-weighted mean velocity
of the 230 GMCs, 
and the rest (``the small clouds", in the following) are listed in Tables 1 and 2, respectively.

The sensitivity of a dataset influences the cloud properties
derived from that data as emphasized by Rosolowsky \& Leroy (2006).
In order to reduce the observational bias,
and to be able to compare datasets with different
signal-to-noise levels,
the boundary of cloud is extrapolated to
a boundary  isosurface of $T_{\rm edge} = 0$ K
(see also Blitz \& Thaddeus 1980; Scoville et al.\ 1987). 
The major and minor axes, size, $R$, position angle, PA, 
and virial mass, $M_{\rm VIR}$
are derived for 164 GMCs
(``Group A GMCs", hereafter)
out of 230; the rest (``Group B", hereafter) have a minor axis less than the NANTEN beam,
so that the size is not derivable by using the de-convolved moment.
The line width, $\Delta V$, and the CO luminosity, $L_{\rm CO}$, 
 are less sensitive to the beam dilution than the cloud size,
so that we can determine $\Delta V$ and $L_{\rm CO}$ of 
the both Group A and B GMCs.
Then, $M_{\rm CO}$, the molecular cloud mass,
is derived from $L_{\rm CO}$ by using 
$X_{\rm CO} = 7 \times 10^{20}$ cm$^{-2}$ (K km s$^{-1}$)$^{-1}$
as derived in Section 4.2 by assuming an virial equilibrium,
and the mass fraction of helium to be 36 \%.
The procedure to derive these properties is described by 
Rosolowsky \& Leroy (2006) in detail. The derived properties are presented in Table 3.

\subsection{Properties of the  Molecular Clouds}

In this section, we shall consider the Group A GMCs, which are
resolved by NANTEN. 
The Group A GMCs have radii ranging from 10 to 220 pc, line widths between 1.6 and 20.2 km s$^{-1}$, CO luminosity between $1.4 \times 10^{3}$ and $7.1 \times 10^{5}$ K km s$^{-1}$ pc$^{2}$, and virial masses ranging from $9 \times 10^{3}$ to $9 \times 10^{6} M_{\odot}$. 

Figure \ref{fig:vlsrhist} shows the frequency distribution of the $V_{\rm LSR}$ of the clouds.
The distribution is rather nonuniform having peaks at $\sim$220--250 km s$^{-1}$ and $\sim$ 280--290 km s$^{-1}$. The one at $V_{\rm LSR} \sim$ 220--250 km s$^{-1}$ represents the clouds in the southern part of the LMC including the molecular ridge and the CO Arc as seen in Figures \ref{fig:ch1}c--e. The other at $V_{\rm LSR} \sim$ 280--290 km s$^{-1}$ is dominated by the emission from the LMC 4, N44, and N11 regions (Figures \ref{fig:ch1}i--j).

Figures \ref{fig:manmin} and \ref{fig:pa164} are the histograms 
of the ratio of the major and minor axes and the position angle (PA) of the GMCs,
respectively. The frequency distribution of PA is rather uniform.
The distribution of the ratio of the major and minor axes has a peak at $\sim 1.7$
with an average of 2.5, indicating that cloud are generally elongated.

In order to see if the cloud has an alignment with the large-scale structure
of the galaxy, such as a spiral pattern,
here we introduce a parameter, $\theta$, 
the angle between the major axis and the tangent at the molecular cloud 
of a circle with a radius, $d_{\rm cen}$ (Figure \ref{fig:theta}a).
Figure \ref{fig:theta}b shows the frequency distribution of $\theta$.
Since the uncertainties of the position angle as well as $\theta$ depend
on the ratio of the major and minor axes,
the histogram of $\theta$ is divided into three groups
according to the axial ratio.
The histogram of $\theta$ shows the number distribution
is nearly uniform and do not have any particular favorable angles.
To see if the distribution of $\theta$ has any characteristics
with the galacto-centric distance, 
Figure \ref{fig:theta}c shows a plot of $d_{\rm cen}$ versus $\theta$.
Again, the plots are distributed quite uniformly
showing no strong dependence of $\theta$ on $d_{\rm cen}$.

\subsubsection{Line-width - size relation}

In this section, we present the correlation between the line width and the radius of the GMCs. 

The size, $R$, of the cloud is computed as a
geometric mean of the ``de-convolved'' second spatial moments along the
major and minor axes which were derived by using the principal component analysis:  we ``de-convolve'' the beam by subtracting its size from the measured cloud size in quadrature.
A full width at half maximum (FWHM) line width, $\Delta V$, is derived by multiplying 
the moment of the velocity within a GMC by $\sqrt{8ln(2)}$
(see also Rosolovsky \& Lorey 2006). 
Figure \ref{fig:rv} shows a plot of log ($\Delta V$) versus log ($R$) of the Group A GMCs. 
It has been known that the line width and the size of molecular clouds have a good correlation of $\sigma_v \propto R^{\sim 0.5}$ in the solar vicinity (Larson 1981) and in the inner Galaxy (e.g., Dame et al.\ 1986; Solomon et al.\ 1987), while the correlation in the GMCs in this work is rather weak, and the best fitting power law 
is $\Delta V = 1.3 R ^{0.2}$.with Spearman rank coefficient of 0.3.
It is likely that the lack of the dynamic range in size may make 
the correlation lower in the present study than for the Galactic clouds, 
although we see a weak positive-correlation in the LMC GMCs.

We have applied the cloud identifying algorithm by Rosolowsky \& Leroy (2006)
with the  parameters used in the current study to identify the GMCs in the LMC
to the molecular clouds in the Small Magellanic Cloud (SMC, Mizuno et al.\ 2001a)
as well as in the outer Galaxy, the Warp region (Nakagawa et al.\ 2005)
and derived the physical properties.
 When we add the clouds in the SMC and the Warp region to the plot of Figure \ref{fig:rv}a, 
the dynamic range in $R$ becomes larger and 
the positive correlation is seen.
A line width-size relation $\sigma_v \propto R^{0.5}$ 
in the inner Galaxy by Solomon et al.\ (1987) as an example
 is also shown as a dotted line in Figure \ref{fig:rv}.
The correlation between the line width and size of the clouds in the LMC, SMC and the Warp region does seem to be consistent with a power law relation $\sigma_v \propto R^{0.5}$ 
but with a clear offset from the relation determined for the inner Galaxy (Solomon et al.\ 1987).  
One has to note that at least a part of this offset can be attributed to differences in the methods used to measure cloud properties. 
The sense of the offset is that for a given radius, the clouds in the
inner Galaxy have larger line widths. This may be partially due to the
relatively high value of $T_A$ used by Solomon et al.~(1987) to
define the cloud radius, implying that the clouds might be smaller for
a given value of $\Delta V$.

\subsubsection{Virial Mass - CO Luminosity Relation}

The determination of the mass of molecular-hydrogen gas is fundamental for understanding the physics of the interstellar medium and star formation in galaxies. In this subsection, we compare the virial mass and the CO luminosity, and discuss 
the conversion factor from the CO line intensity to the H$_{2}$ column density 
in the LMC.

Figure \ref{fig:mvl} shows the virial mass, $M_{\rm VIR}$, as a function of  luminosity, $L_{\rm CO}$, of the clouds in the LMC (filled circle). The plot shows a tight power law for the mass luminosity relation with some dispersion. A least-squares fit to the data gives a power law, 
[$M_{\rm vir}$/$M_{\odot}$]= 
26[$L_{\rm CO}$/(K km s$^{-1}$ pc$^{2}$)]$^{1.1\pm0.3}$,
with Spearman rank correlation of 0.8. 
This relation suggests that clouds are virialized and CO luminosity can be a good tracer of mass in the LMC with a quite constant conversion factor from $L_{\rm CO}$ to mass throughout the mass range $10^4 \la M_{\rm VIR} / M_{\sun} \la 10^{7}$. 
We have added the re-identified clouds in the SMC (Mizuno et al. 2001a) and 
the Warp region (Nakagawa et al.\ 2005) to Figure \ref{fig:mvl} as in Section 3.3.1. 
The clouds in the SMC and the Warp region lie along the best fitting power-law of GMCs in the LMC.

\subsubsection{Mass Spectrum}

The frequency distribution of the cloud masses has an important impact
not only in the star formation but also in cloud formation and destruction.
The mass spectrum of the clouds is well fitted by a power law and
often presented as $dN/dM \propto M^{-(\alpha +1)}$ or
$N_{\rm cloud}$ ($> M$) $\propto M^{-\alpha}$.
The preliminary results of the mass spectrum by the second NANTEN survey in the LMC
have been already presented and discussed in Fukui et al.\ (2001).
They found the mass spectrum derived from 
the CO luminosity has a slope with $\alpha = 0.9 \pm 0.1$
above the completeness limit of $8 \times 10^4 \, M_{\odot}$.

In this section, we present the mass spectra of the mass, $M_{\rm CO}$, 
derived from a CO luminosity, $L_{\rm CO}$, and  
a conversion factor, $X_{\rm CO} = 7 \times 10^{20}$ cm$^{-2}$ K km s$^{-1}$
 (section 3.2 and section 4.2).
Here, the CO luminosity, $L_{\rm CO}$, is less sensitive to the beam dilution than the cloud size,
so that we can consider that not only the Groups A GMCs but also
the $L_{\rm CO}$ of the Group B GMCs,
which we could not derive a size and $M_{\rm VIR}$,
are determined well enough to obtain $M_{\rm CO}$ (Section 3.2 and Table 3).
The mass spectrum of $M_{\rm CO}$  including both the Group A and B GMCs,
is shown in Figure \ref{fig:ms}.
The maximum likelihood method (Crawford et al.\ 1970) was applied 
to obtain the best-fitting power law above the completeness limit, $5 \times 10^4 M{\odot}$.
The best fitting power law above the completeness limit  is
$N_{\rm cloud}$($\geq M_{\rm CO}$) = $6.6 \times 10^{5} M^{-0.75 \pm 0.06}-3.4$.

The results indicate that mass of the molecular gas in the LMC is 
concentrated in the massive clouds, since $\alpha < 2$. 
The slope of the mass spectrum, thus the fact that the massive clouds
contribute to the galactic total mass, is consistent within the current results
as well as with what is presented by Fukui et al.\ (2001), 
although the current result shows shallower slope than the other.
This difference in the index values of the best-fitting power law may be explained by 
the difference in completeness limit.
Table 4 shows the index value, $\alpha$, of the power law
fit to the different mass range of the GMCs.
The best fitting power law obtained for the clouds with 
$M_{\rm CO} \geq 3 \times 10^{5} M_{\sun}$
is shown in Figure \ref{fig:ms} as an example.
Table 4 indicates that 
the slope of the mass spectrum becomes steeper if we fit only the massive
clouds; e.g., $N_{\rm cloud}$($>M_{\rm CO}$) $\propto M_{\rm CO}^{-1.2 \pm 0.2}$ for $M_{\rm CO} \ge 3 \times 10^{5} M_{\sun}$,
and the logarithmic slope of the mass spectrum 
becomes steeper at $\sim 3 \times 10^5 M_{\sun}$.

\section{Discussion}
\subsection{Comparison with the first Survey}
The 2nd survey was carried out to cover the regions where the molecular clouds are detected
in the 1st survey. 
The signal-to-noise ratio of the present observations was higher by a
factor of 2 than that in the 1st NANTEN survey (Fukui et al.\
1999; Mizuno et al.\ 2001b; Yamaguchi et al.\ 2001c).
This increase in sensitivity made possible to
increase the number of the significant detections and identified clouds.
Different cloud identification criteria are used in the current study and
those used in Fukui et al. (1999) and Mizuno et al. (2001b). 
Nevertheless, the number of the clouds with more than two observing positions
(``the large clouds in Mizuno et al. (2001b) ) 
is a factor of 3 larger in the present survey even
if we use the same algorithm and criterion.

Here we compare the line width-size relation and 
virial mass-CO luminosity relation of the molecular clouds 
from the 1st and the current surveys.
To compare the results, we re-calculated the size and the virial mass 
of the large clouds of the 1st survey
by subtracting the NANTEN beam from the size in Table 1 of Mizuno et al.\ (2001b).

Figure \ref{fig:rdv1st} is a plot of the line width and the size of the clouds
derived from both the 1st survey (open circle) and the current survey (filled circle).
The clouds from both the current and the 1st survey show a
large scatter with a little positive correlation of the line width and the size.
An offset from the correlation of the inner Galaxy is also seen.
The scatter is larger in the current survey.
This may be explained by the difference in the sensitivity;
the high sensitivity of the current survey made it possible to decompose 
a cloud into several individual clouds with different velocities along the same line of site, 
although these clouds may have been identified as an entity by the 1st survey.

Figure \ref{fig:mvlco1st}  is a plot of the virial mass, $M_{\rm VIR}$
as a function of luminosity, $L_{\rm CO}$, of the Group A GMCs from 
the current survey (filled circle) and the 55 clouds from the 1st survey (open circle).
The correlations between the $M_{\rm VIR}$ and $L_{\rm CO}$ in both surveys 
are consistent within the error.
Again, the scatter in the current survey is larger,
but because the higher sensitivity limit of the current survey enlarges 
the dynamic range in $M_{\rm VIR}$ and $L_{\rm CO}$ 
the correlation coefficient remains as high as 0.85.
The ratio of the $M_{\rm VIR}$ and $L_{\rm CO}$ is 
related to the conversion factor, $X_{\rm CO}$-factor, 
from the CO luminosity to the hydrogen column density, $N$(H$_{\rm 2}$).
The consistency of the $M_{\rm VIR}$ and $L_{\rm CO}$ relation
in both survey suggests that the $X_{\rm CO}$-factor derived
from the both surveys are also consistent.

\subsection{CO to $N$($H_2$) Conversion Factor}

In the following, we derive $X_{\rm CO}$-factor from the current survey.

The determination of the mass of H$_2$ in galaxies is fundamental for an understanding of the interstellar physics and star formation. The principal method for obtaining H$_2$ masses converts the intensity of the CO molecular line emission, $I_{\rm CO}$, into the column density of H$_2$ molecules. The conversion factor, 
$X$-factor ($X \equiv N$(H$_{2}$) $/$ $I_{\rm CO} = M_{\rm H_2}/L_{\rm CO}$), 
has been derived for molecular clouds in the solar vicinity based on the assumption of virial equilibrium of individual clouds (e.g., Young \& Scoville 1991). 
This method has been also used to derive the conversion factors in nearby galaxies,
such as the LMC, SMC, M31, M33 and etc., where individual clouds are resolved
(e.g., Mizuno et al. 2001a; Mizuno et al. 2001b; Wilson \& Scoville 1990).

The plot of the $M_{\rm VIR}$ against $L_{\rm CO}$ in section 3.3.2
(Figure \ref{fig:mvl}) from the current survey suggests that 
for massive clouds, $M_{\rm CO} > 10^5 M_{\sun}$, it is reasonable to assume
virial equilibrium.
Here, we shall use the conventional method, by applying the virial theorem to 
the clouds to estimate the $X_{\rm CO}$-factor in the LMC.
The average value of log($M_{\rm VIR} / L_{\rm CO}$) is $1.2 \pm 0.3$, 
corresponding to $X_{\rm CO} = (7 \pm 2) \times 10^{20}$ cm$^{-2} $(K km s$^{-1}$)$^{-1}$,
 for the Group A GMCs with
$L_{\rm CO}$ higher than the completeness limit, 
$9 \times 10^{3}$ K km s$^{-1}$ pc$^{2}$, 
that is equivalent to $M_{\rm VIR} \geq 1.4 \times 10^{5} M_{\odot}$.
Figure \ref{fig:mvlhist} is a frequency distribution of log($M_{\rm VIR} / L_{\rm CO}$)
of the Group A GMCs, 
i.e., including all the clouds for which we derived virial masses.
The geometric mean of $M_{\rm VIR} / L_{\rm CO}$, and thus the
$X_{\rm CO}$-factor, do not differ from the values obtained
by using the clouds with 
$L_{\rm CO} \ge 9 \times 10^{3}$ K km s$^{-1}$ pc$^{2}$ only.

This value is slightly less than what we obtained from the 1st survey,
$X_{\rm CO} \sim (8 \pm 2) \times 10^{20}$cm$^{-2}$(K km s$^{-1}$)$^{-1}$,
 after taking into account the beam de-convolution,
 but is consistent within the error.

An $X_{\rm CO}$-factor in the inner Galaxy has been derived by using a correlation between 
the $\gamma$ ray intensity and the CO intensity along the Galactic plane (Bloemen et al.\ 1986).
Bloemen et al.\ (1986) summarized the value  of $X$ factors and the value for the inner Galaxy is derived to be $\sim$ (1--3)$\times 10^{20}$cm$^{-2}  $(K km s$^{-1}$)$^{-1}$ 
on average.
The $X$-factor obtained above is about twice higher than that of the clouds in the inner Galaxy.

Bertoldi \& McKee (1992) argue that the gravitational energy, $W$, 
of an ellipsoidal cloud is given by
$W = -3/5 \times [GM^2 / R] \times [arcsin(e)/e]$,
(Eq. A9 of Bertoldi \& McKee 1992) where $G$, $M$, and $R$ are the gravitational constant, 
mass and the size of the clouds. Here, $e$ is an eccentricity of the cloud, $e = (1-y^2)^{1/2}$, 
where y is an axial ratio of the cloud.
The current sample of the GMCs are not really spherical with 
the mode of the ratio of the major and minor axes to be $\sim 1.7$
and the average 2.5 as shown in section 3.3 (see also Figure 8).
The shape-dependent factor, $a_{2} = R_{m} /R$ $arcsin(e)/e$ (Eq. A8 of Bertoldi \& McKee 1992), 
ranges from 1 to 0.88 for the current sample of the GMCs in the LMC with $a_2 = 0.99$
for the axis ratio of 2.5.
This argument means that the virial mass of the current sample of the GMCs 
differs from the derived virial mass with about 15 \% at most due to the elliptical shape of the clouds.
The deviations of the estimated $M_{\rm VIR}$ from the true $M_{\rm VIR}$
 affect the derived $X_{\rm CO}$ factor linearly.
Thus the current estimate of the $X_{\rm CO}$ factor can be overestimated by $\sim15 \%$ at most
due to the discrepancy of the cloud shape from the spherical symmetry.

Rubio et al.\ (1993) suggested a possible dependence of the $X_{\rm CO}$-factor
on the cloud size from their observation toward the clouds in the SMC
by SEST. Figure \ref{fig:mvl_r} is a plot of the $M_{\rm VIR}/L_{\rm CO}$
against the cloud size. There is no significant correlation in the LMC clouds
with scatter as large as an order of magnitude in $M_{\rm VIR}/L_{\rm CO}$,
although the range of the cloud sizes is the same as that of the SMC (Rubio et al.\ 1993)

It has been claimed that the metallicity is quite uniform in the LMC (e.g., Dufour 1984), while it has been suggested that
the metallicity of the outer part of a galaxy is lower than that of the inner region in the Galaxy as well as in some of the nearby galaxies (e.g., Nakagawa et al. 2005).
Figure \ref{fig:dcen_lco} shows a plot of 
$M_{\rm VIR}/L_{\rm CO}$ against the distance from the center of the LMC derived from the HI distribution (Kim et al.\ 1998).
The current result shows no clear correlation of 
$M_{\rm VIR}/L_{\rm CO}$ to the distance from the center,
suggesting that $X_{\rm CO}$-factor does not depend on the distance from the center of the LMC. This is consistent with the idea that the metallicity is quite uniform in the LMC.

\subsection{Mass Spectrum}

The mass spectra of the Galactic clouds ($^{12}$CO or$^{13}$CO) are well fitted by a power law with index values of $\alpha$ $\sim$0.5 -- 1.0 (e.g., Solomon et al.\ 1987; Solomon \& Rivolo 1989; Casoli, Combes, \& Gerin 1984;  Digel et al.\ 1996; Dobashi et al.\ 1996),
of which the higher values are derived for the clouds with lower mass in the outer Galaxy
(e.g., Heyer et al.\ 2001) or those derived by using the data from $^{13}$CO observations
(e.g., Yonekura et al.\ 1997; Kawamura et al.\ 1998).
Not only the mass distribution of the Galactic clouds but also that of the clouds in nearby galaxies have been studied (Blitz et al. 2006 and the references therein). 
Most of the galaxies have similar mass distributions, $\alpha \sim 0.7$; 
an exception is M33, which is steeper than the rest but the dataset has a higher completeness limit (Rosolowsky et al.\ 2003; Rosolowsky et al.\ 2005).

A number of numerical simulations have been conducted to obtain
a mass spectrum of GMCs in galaxies 
(e.g., V\'{a}quez-Semadeni et al.\ 1997; Wada et al.\ 2000, and reference therein). 
Wada et al.\ (2000) carried out a simulation to 
the H {\sc i} and CO distributions of the LMC-type galaxy  
specifically at the highest spatial resolution ($\sim$ 7.8 pc) by 
incorporating fairly realistic star-formation processes and 
supernova rates. 
They show that the GMC mass spectrum in an LMC-like galaxy 
is expected to have a power law with an index value of $\sim 0.7$
 if no star formation is taken into account, and   
that the index becomes steeper around $1.0$ if the
dissipation of clouds due to star formation is incorporated.
The present mass spectrum appears to be consistent with these values, 
while those of the non star-forming models may fit slightly better the present result. 
According to Wada \& Norman (2001), the absence of a massive GMC of 
$\sim 3 \times 10^{6} \, M_{\Sol}$ is due to the well-mixed, turbulent interstellar medium 
that tends to form smaller GMCs. 
The consistency of the mass spectrum among galaxies may suggest that the mechanism of cloud formation and the disruption show similar characteristics among the galaxies.
Nevertheless, the truncation of very massive GMCs 
may suggest that the disruption of the molecular clouds 
is faster in the massive clouds.
It may also suggest that cloud formation takes place inhomogeneously;
the mass spectra in different regions of the galaxy may have different slopes
and the truncation of the slope might appear
when we sum up all the mass spectra within the galaxy.
The reason of the truncation is not yet know but 
the current result present new information leading to a better knowledge
of the cloud formation and disruption.

\section{Summary} \label{summary}
A large-scale $^{12}$CO($J$ = 1--0) survey for the molecular
clouds was made toward the Large Magellanic Cloud by NANTEN.
An area of $\sim$ 30 square degrees was covered, and 
significant $^{12}$CO emission
($\geq$ 0.07 K ) was detected at $\sim 1,300$
out of the 26,900 observed positions.
We identified 272 molecular clouds, 230 of which were detected at
more than three observing positions.
The position, mean velocity and velocity dispersion of the 272 clouds,
and the extents, position angle and CO luminosity of the 230 GMCs are derived. 
A reliable size and virial masses are determined for the well resolved 164 GMCs
(Group A).
The main results are summarized as follows.

\begin{enumerate}
\item The Group A GMCs have radii ranging from 10 to 220 pc, 
line widths between 1.6 and 20.2 km s$^{-1}$, 
CO luminosities between $1.4 \times 10^{3}$ and 
$7.1 \times 10^{5}$ K km s$^{-1}$ pc$^{2}$, 
masses derived from CO luminosity from 
$2 \times 10^{4}$ to $7 \times 10^{6} M_{\sun}$,
and virial masses from 
$9 \times 10^{3}$ to $9 \times 10^{6} M_{\odot}$. 
The maximum temperature ($T_{\rm R}^{*}$) 
of the CO line is as high as $\sim$ 2 K, 
which is detected toward the N 113 and N 159 regions. 

\item The line width, $\Delta V$, 
and the radius, $R$, 
of the Group A GMCs appear to satisfy the slope of the power-law in line width-size
relation of the clouds in the Galaxy, but with an offset in the constant of
proportionality. 

\item A least-squares fit to virial mass vs. CO luminosity relation
shows a power law, [$M_{\rm vir}$/$M_{\odot}$]= 
26[$L_{\rm CO}$/(K km s$^{-1}$ pc$^{2}$)]$^{1.1\pm0.3}$
with Spearman rank correlation of 0.8.  
This good correlation shows 
that the CO luminosity is   
a good tracer of the mass of molecular clouds in the LMC. 

\item The $I_{\rm CO}$--$N$(H$_{2}$) conversion factor 
is found to be 
$X_{\rm CO} \sim$
$7 \times 10^{20}$cm$^{-2} $(K km s$^{-1}$)$^{-1}$
by assuming the virial equilibrium from the Group A GMCs.

\item The mass spectrum of the GMCs with $
5 \times {10}^4 \le M_{\rm CO} \le {10}^7 M_{\sun}$
is well fitted by a power
law, $N_{\rm CO}$($>M_{\rm CO}) \propto {(M_{\rm
CO}/M_{\sun})}^{-0.75 \pm 0.06}$. This slope is consistent with the previous
results obtained from the Galaxy and nearby galaxies.
The slope of the mass spectrum becomes steeper if we fit only the massive
clouds; e.g., $N_{\rm cloud}$($>M_{\rm CO}$) $\propto M_{\rm CO}^{-1.2 \pm 0.2}$ for $M_{\rm CO} \ge 3 \times 10^{5} M_{\sun}$,
suggesting the mass truncation.

\end{enumerate}

\acknowledgments
The NANTEN project is based on a mutual agreement between Nagoya University and the Carnegie Institution of Washington (CIW). We greatly appreciate the hospitality of all the staff members of the Las Campanas Observatory of CIW. 
We acknowledge Drs Blitz, Rosolowsky and Leroy for their discussion and providing us with their cloud identifying program.
We are thankful to many Japanese public donors and companies who contributed to the realization of the project. This work is financially supported in part by a Grant-in-Aid for Scientific Research from the Ministry of Education, Culture, Sports, Science and Technology of Japan (No.\ 15071203) and from JSPS (No.\ 14102003, core-to-core program 17004 and No.\ 18684003). TM is supported by the Japan Society of the Promotion of Science,
and MR by the Chilean {\sl Center for Astrophysics}
FONDAP No. 15010003.

\clearpage

%\documentclass[preprint]{aastex}
%\begin{document}
\begin{deluxetable}{ccccccc} 
\rotate
\tablecolumns{7} 
\tablewidth{0pc} 
\tabletypesize{\footnotesize}
\tablecaption{Catalog of 230 GMCs} 
\tablehead{ &&\multicolumn{4}{c}{Position\tablenotemark{b}}&
\colhead{Cloud property\tablenotemark{c}}\\
\cline{3-7}\\
\colhead{Number}&\colhead{Name\tablenotemark{a}}&\colhead{$\alpha$(2000)}&\colhead{$\delta$(2000)}&
\colhead{$\alpha$(1950)}&\colhead{$\delta$(1950)}&
\colhead{$V_{\rm LSR}$}\\
&&\colhead{(h m s)}&\colhead{($\arcdeg$ $\arcmin$ $\arcsec$)}&
\colhead{(h m s)}&\colhead{($\arcdeg$ $\arcmin$ $\arcsec$)}&
\colhead{(km s$^{-1}$)}}
\startdata 
1  &  LMC N J0447-6910  &  4  47  30.4  &  -69  10  \phn8.9  &  4  47  44.0  &  -69  15  20.2  &  239.4\\
2  &  LMC N J0447-6713  &  4  47  35.9  &  -67  13  44.0  &  4  47  33.8  &  -67  18  55.4  &  256.8\\
3  &  LMC N J0448-6920  &  4  48  40.8  &  -69  20  41.3  &  4  48  56.2  &  -69  25  47.6  &  238.9\\
4  &  LMC N J0449-6910  &  4  49  12.4  &  -69  10  21.0  &  4  49  26.4  &  -69  15  25.2  &  240.3\\
5  &  LMC N J0449-6826  &  4  49  24.2  &  -68  26  49.8  &  4  49  32.0  &  -68  31  53.4  &  249.4\\
6  &  LMC N J0449-6806  &  4  49  25.0  &  -68  \phn6  39.1  &  4  49  30.0  &  -68  11  42.7  &  251.0\\
7  &  LMC N J0449-6652  &  4  49  33.0  &  -66  52  36.9  &  4  49  28.7  &  -66  57  40.3  &  264.8\\
8  &  LMC N J0450-6910  &  4  50  23.0  &  -69  10  45.3  &  4  50  37.3  &  -69  15  44.6  &  237.8\\
\enddata 
\tablenotetext{a}{Name of the clouds; `N' stands for `NANTEN'.}
\tablenotetext{b}{Position of the clouds derived from an intensity-weighted mean over all the pixels of a cloud.}
\tablenotetext{c}{$V_{\rm LSR}$ is an LSR velocity derived from an intensity-weighted mean of the velocity distribution
within a cloud.}
\end{deluxetable}
%\end{document}

%\documentclass[preprint]{aastex}
%\begin{document}
\begin{deluxetable}{ccccccc} 
\rotate
\tablecolumns{7} 
\tablewidth{0pc}
\tablenum{2} 
\tabletypesize{\footnotesize}
\tablecaption{Catalog of 42 small clouds} 
\tablehead{ & &\multicolumn{4}{c}{Position\tablenotemark{b}}&
\colhead{Cloud property\tablenotemark{c}}\\
\cline{3-7}\\ 
\colhead{Number\tablenotemark{a}}&\colhead{Name}&\colhead{$\alpha$(2000)}&\colhead{$\delta$(2000)}&
\colhead{$\alpha$(1950)}&\colhead{$\delta$(1950)}&
\colhead{$V_{\rm LSR}$}\\
&&\colhead{(h m s)}&\colhead{($\arcdeg$ $\arcmin$ $\arcsec$)}&
\colhead{(h m s)}&\colhead{($\arcdeg$ $\arcmin$ $\arcsec$)}&
\colhead{(km s$^{-1}$)}}
\startdata 
231  &  LMC N J0447-6728  &  4  47  18.3  &  -67  28  47.5  &  4  47  18.0  &  -67  34  \phn0.1  &  250.0\\
232  &  LMC N J0448-6721  &  4  48  \phn1.7  &  -67  21  53.9  &  4  48  \phn0.7  &  -67  27  \phn3.6  &  254.1\\
233  &  LMC N J0449-6810  &  4  49  40.6  &  -68  10  12.9  &  4  49  46.2  &  -68  15  15.5  &  255.2\\
234  &  LMC N J0451-6927  &  4  51  21.8  &  -69  27  \phn4.8  &  4  51  38.8  &  -69  31  59.9  &  237.7\\
235  &  LMC N J0453-6846  &  4  53  55.9  &  -68  46  38.1  &  4  54  \phn7.5  &  -68  51  22.7  &  242.7\\
236  &  LMC N J0458-6617  &  4  58  50.0  &  -66  17  \phn7.1  &  4  58  43.2  &  -66  21  31.7  &  273.8\\
237  &  LMC N J0459-6607  &  4  59  50.1  &  -66  \phn7  39.6  &  4  59  42.4  &  -66  12  \phn0.0  &  279.4\\
238  &  LMC N J0500-6807  &  5  \phn0  36.9  &  -68  \phn7  43.4  &  5  \phn0  44.3  &  -68  12  \phn0.0  &  256.4\\
\enddata 
\tablenotetext{a}{Name of the clouds; `N' stands for `NANTEN'.}
\tablenotetext{b}{Position of the clouds derived from an intensity-weighted mean over all the pixels of a cloud.}
\tablenotetext{c}{$V_{\rm LSR}$ is an LSR velocity derived from an intensity-weighted mean of the velocity distribution
within a cloud.}
\end{deluxetable}
%\end{document}

\clearpage
%\setlength{\hoffset}{-20mm}
%\documentclass[preprint]{aastex}
%\begin{document}
\begin{deluxetable}{ccccccccc} 
\rotate
\tablenum{3}
\tablecolumns{9} 
\tablewidth{0pc} 
\tabletypesize{\footnotesize}
\tablecaption{Properties of 230 GMCs} 
\tablehead{
& &
\multicolumn{7}{c}{Cloud properties\tablenotemark{a}}\\
\cline{3-9}\\ 
\colhead{Number}&
\colhead{Name}&
\colhead{$\Delta V_{\rm LSR}$}&
\colhead{$R$\tablenotemark{b}}&
\colhead{PA}&
\colhead{Maj$_{\rm nodeconv}$ $\times$ Min$_{\rm nodeonv}$\tablenotemark{c}}&
\colhead{$L_{\rm CO}$}&
\colhead{$M_{\rm vir}$\tablenotemark{b}}&
\colhead{$M_{\rm CO}$}\\
 & &
\colhead{(km s$^{-1}$)}&
\colhead{(pc)}&
\colhead{($\arcdeg$)}&
\colhead{(pc \phn pc)}&
\colhead{($\times 10^{4}$ K km s$^{-1}$)}&
\colhead{($\times 10^{5} M_{\sun}$)}&
\colhead{($\times 10^{5} M_{\sun}$)}}
\startdata 
1  &  LMC N J0447-6910  &  \phn4.2  &  \nodata  &  \phn46  &  \phn32  $\times$  15   &  \phn1  &  \nodata  &  \phn2\\
2  &  LMC N J0447-6713  &  11.2   &  \phn97  &  \phn54  &  \phn69  $\times$  42   &  \phn7  &  \phn20  &  \phn10\\
3  &  LMC N J0448-6920  &  \phn4.1  &  \nodata  &  169   &  \phn39  $\times$  14   &  \phn0.5  &  \nodata  &  \phn0.7\\
4  &  LMC N J0449-6910  &  \phn5.3  &  \phn60  &  \phn23  &  \phn51  $\times$  26   &  \phn3  &  \phn3  &  \phn4\\
5  &  LMC N J0449-6826  &  \phn5.7  &  \phn87  &  \phn87  &  \phn79  $\times$  31   &  \phn8  &  \phn5  &  \phn10\\
6  &  LMC N J0449-6806  &  10.7   &  \nodata  &  \phn73  &  \phn32  $\times$  13   &  \phn0.9  &  \nodata  &  \phn1\\
7  &  LMC N J0449-6652  &  \phn4.6  &  \phn26  &  \phn59  &  \phn23  $\times$  20   &  \phn0.5  &  \phn1  &  \phn0.9\\
8  &  LMC N J0450-6910  &  \phn1.9  &  \nodata  &  179   &  \phn30  $\times$  12   &  \phn0.4  &  \nodata  &  \phn0.6\\
\enddata
\tablenotetext{a}{Observed properties of the GMCs identified from this survey;
FWHM of the line width, $\Delta V$, derived from the moment of the spectra within a cloud,
size, $R$, derived by $R = 1.91 \times \sqrt{Maj^2 \times Min^2}$, where $Maj$ and $Min$ are
the major and minor axes of the cloud deconvolved by the NANTEN beam, respecticely.
Position angle, major and minor axes without beam deconvolution, CO luminosity, $L_{\rm CO}$, 
virial mass, $M_{\rm VIR}$, and mass, $M_{\rm CO}$, derived from $L_{\rm CO}$ with 
$X_{\rm CO} = 7 \times 10^{20}$ cm${^-2}$ (K km s$^{-1}$)$^-1$  are also listed.
Please also see section 3.2 for the derivation of the cloud properties.}
\tablenotetext{b}{Minor axis of clouds less than the NANTEN beam is not resolved
by NANTEN and is not able to be derived. The radius and the virial mass of these clouds
are indicated by `\nodata'.}
\tablenotetext{c}{Minor axis without deconvolution indicated by `\nodata' is that of a cloud only detected at one observed point
along the minor axis so that the moment is not derivable.}
\end{deluxetable} 

% \end{document} 
\clearpage
%\setlength{\hoffset}{0mm}
%\documentclass[preprint]{aastex}
%\begin{document}
\begin{deluxetable}{ccc} 
\tablecolumns{3} 
\tablewidth{0pc}
\tablenum{4} 
\tablecaption{Slope of mass spectrum} 
\tablehead{
\colhead{Mass range\tablenotemark{a}} & 
\colhead{$\alpha$\tablenotemark{b}} & 
\colhead{Number of clouds\tablenotemark{c}}\\
\colhead{$\times 10^{4} M_{\odot}$} & & 
}
\startdata 
\phn$5<$& 0.75 $\pm$ 0.06 & 191\\
%\phn$6<$& 0.83 $\pm$ 0.07 & 178\\
%\phn$7<$& 0.87 $\pm$ 0.08 & 162\\
\phn$8<$& 0.88 $\pm$ 0.08 & 146\\
%\phn$9<$& 0.89 $\pm$ 0.09 & 132\\
$10<$& 0.93 $\pm$ 0.10  & 123\\
$13<$& 0.98 $\pm$ 0.11 & 100\\
%$14<$& 0.99 $\pm$ 0.12 & 94\\
%$15<$& 0.98 $\pm$ 0.12 & 86\\
$20<$& 1.0 $\pm$ 0.1 & 67\\
$30<$& 1.2 $\pm$ 0.2 & 47\\
%$40<$& 1.4 $\pm$ 0.3 & 37\\
$50<$& 1.6 $\pm$ 0.3 & 28\\
\enddata 
\tablenotetext{a}{Clouds from the 230 GMCs with mass 
indicated in this column are used to derive a mass spectrum.}
\tablenotetext{b}{Index value of the best-fitting power law,
$N_{\rm cloud}$ ($> M$) $\propto M^{-\alpha}$,
to the mass spectrum of the clouds with mass indicated in the 1st column.}
\tablenotetext{c}{Number of the clouds used to fit the power law.}
\label{tab:ms}
\end{deluxetable}
%\end{document}

\clearpage

\begin{figure}
\epsscale{0.6}
\plotone{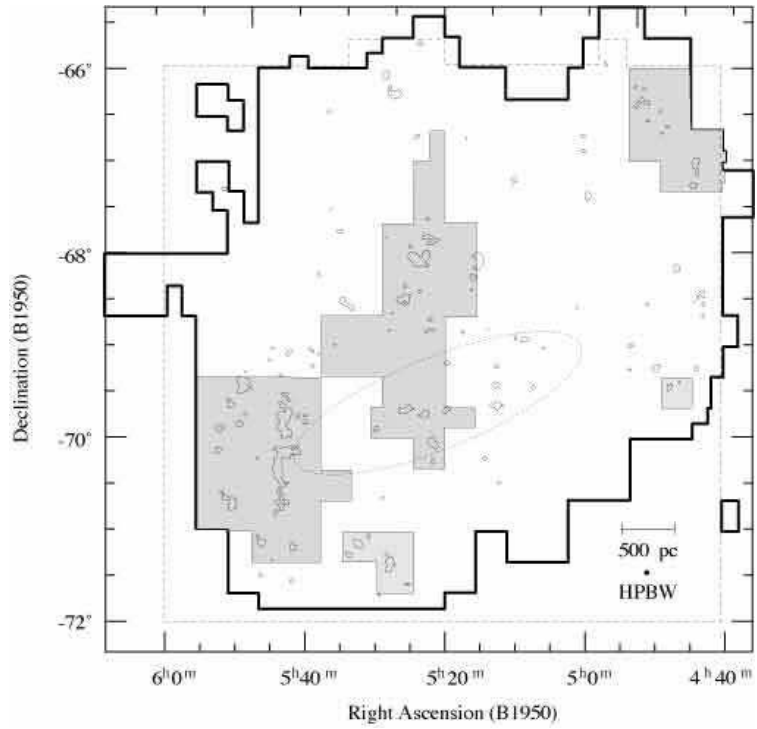}
\caption{
The boundary of the observed region in this work is indicated by the
thick sold lines;
the areas with gray
indicate the regions observed by the narrow band spectrometer
(see section 2).
Overlaid is the velocity-integrated intensity distribution of CO of the 1st survey,
whose boundary is shown by the dashed line
(Fukui et al.\ 1999; Mizuno et al.\ 2001b).
The contours are at $5 \sigma$ noise level (3 K km s$^{-1}$) of the velocity-integrated
intensity of the 1st survey, showing that 
the present survey covers the entire regions where CO emission
were detected. 
The ellipse illustrates the position of the bar.
}
\label{fig:obsreg}
\end{figure}
\clearpage

\begin{figure*}
\epsscale{1.0}
\plotone{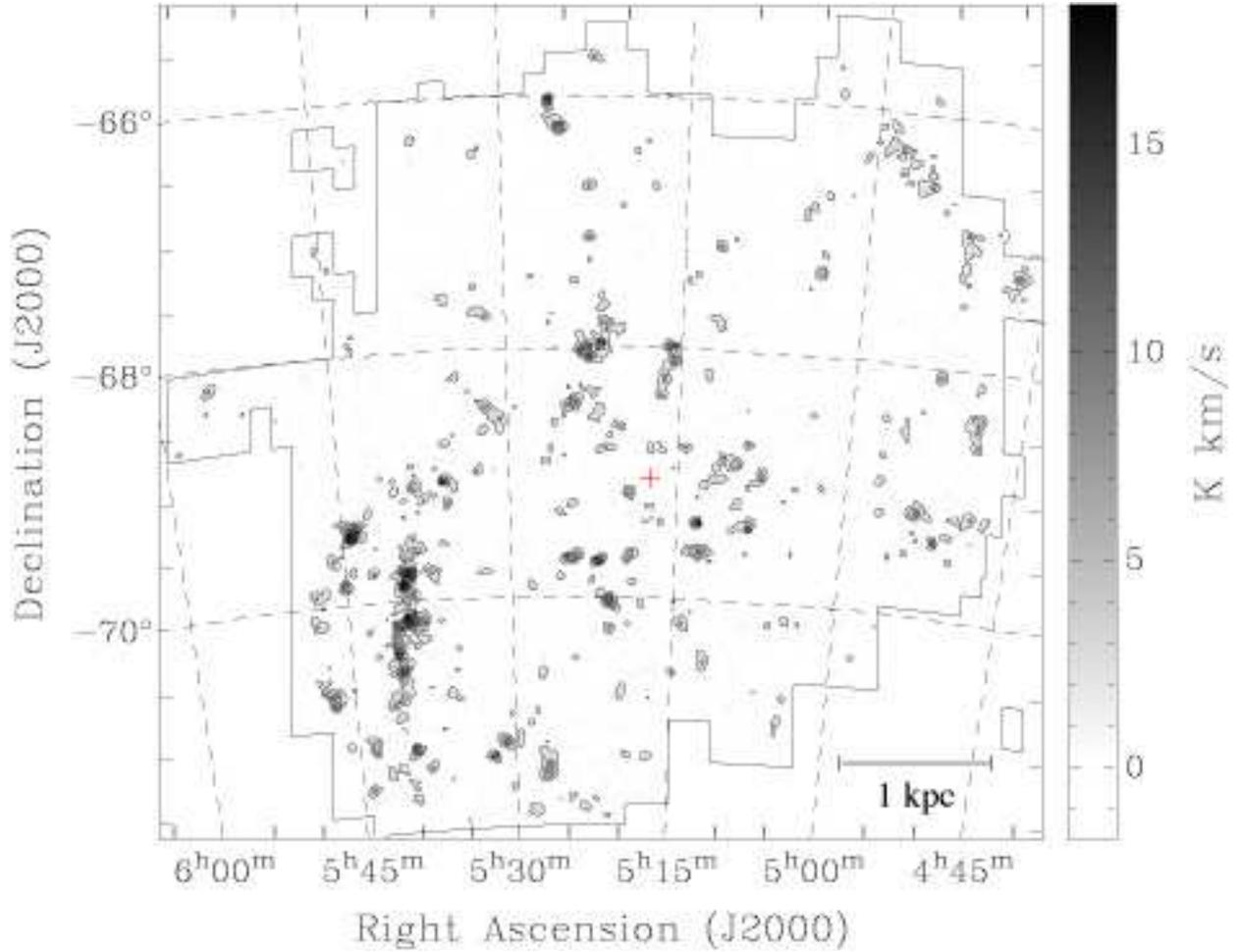}
\caption{(a) Velocity-integrated map of the $^{12}$CO emission integrated over
$150 \le V_{\rm LSR} < 350$ km s$^{-1}$.
The contours are at 1.2 ($3\sigma$ noise level), 3.6, 6.0, 8.4, 12.0, 15.6 K km s$^{-1}$. 
The thin lines show the observed region and 
the cross indicate the center determined from the kinematics of the HI observations by Kim et al.\ (1998).
(b) Velocity-integrated map of the $^{12}$CO emission
superposed on an optical image.
The contours are the three lowest ones from 1.2 K km s$^{-1}$ with 
2.4 K km s$^{-1}$ intervals,
while only the lowest three contours are shown not to mask the optical features.
}
\label{fig:ii}
\end{figure*}
\clearpage

\begin{figure*}
\epsscale{1.0}
{\plotone{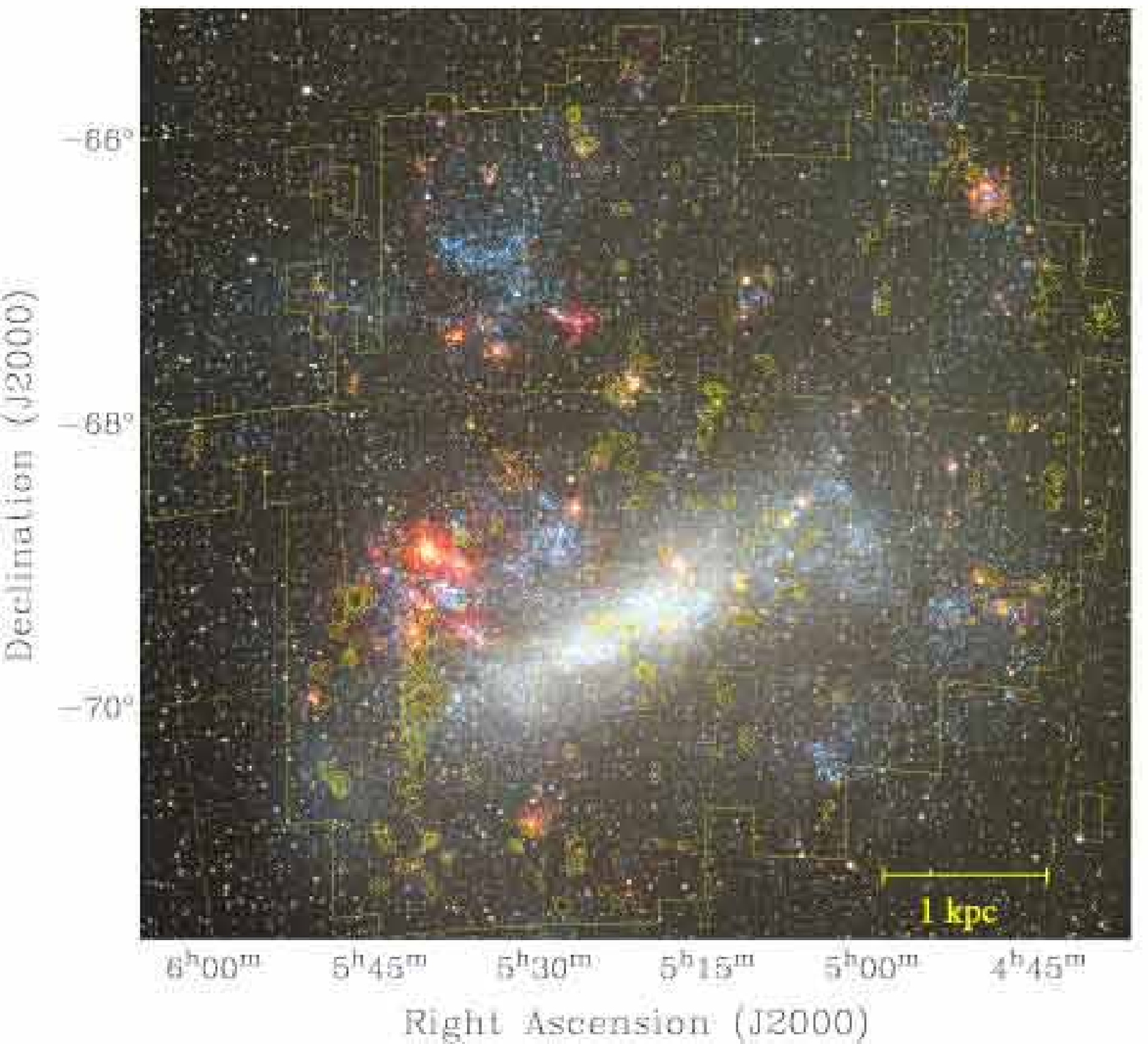}\\[5mm]}
\centerline{Fig. 2. --- {\it continued}}
\label{fig:ii_col}
\end{figure*}
\clearpage

\clearpage
\begin{figure}
\epsscale{0.6}
\plotone{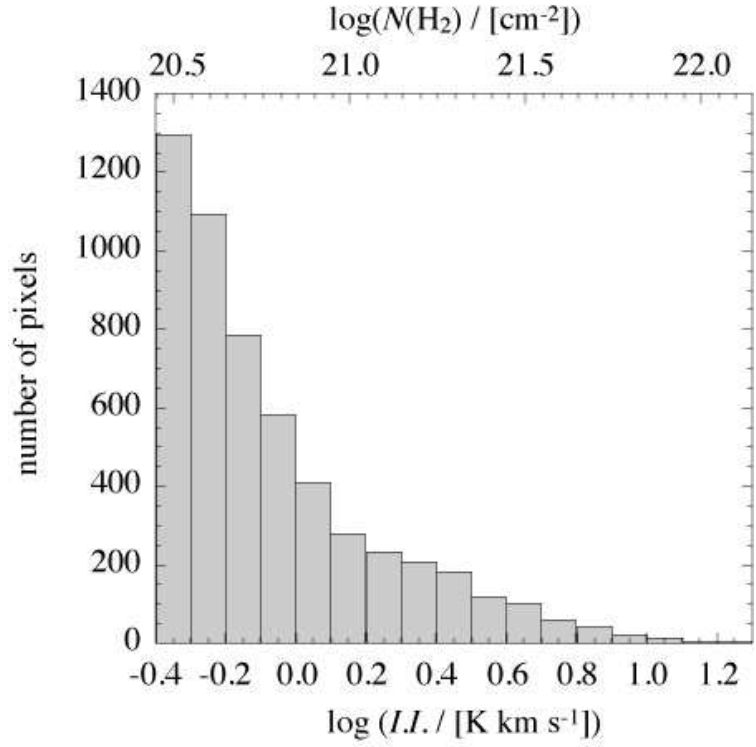}
\caption{Distribution of the integrated intensity greater than 0.4 K km s$^{-1}$ ($\sim 1\sigma$ noise level) of each observing position.  The equivalent values of the column density 
derived from the integrated intensity by using 
$X_{\rm CO} = 7 \times 10^{20}$ cm$^{-2}$ (K km s$^{-1}$)$^{-1}$
are also shown.
}
\label{fig:iihist}
\end{figure}

\clearpage
\begin{figure}
\epsscale{0.5}
\plotone{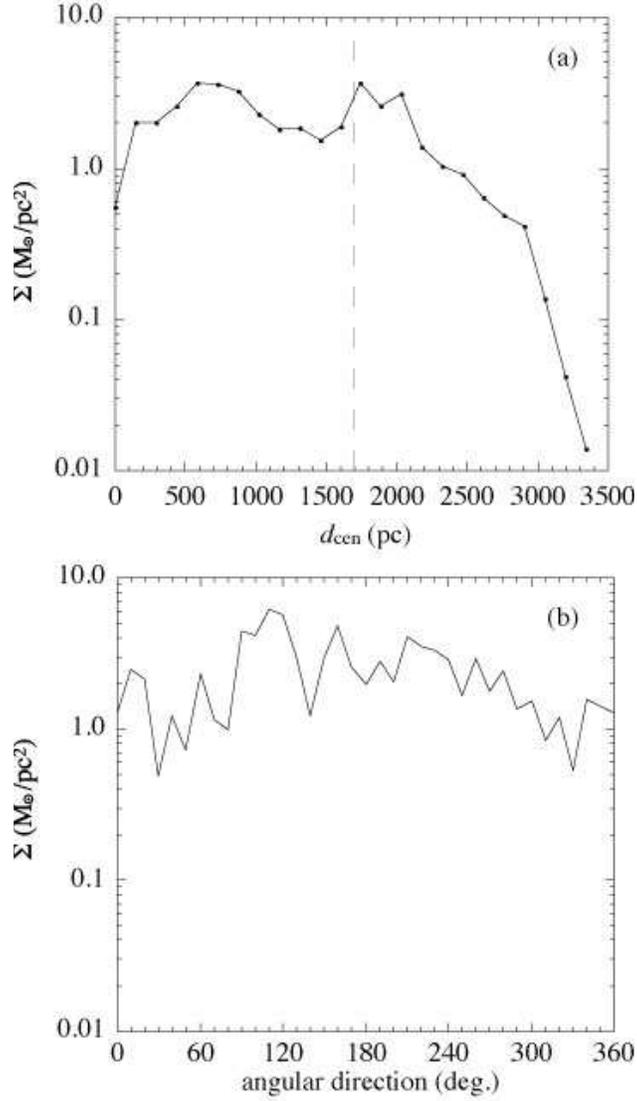}
\caption{(a) Distribution of the surface density along the distance from the center
$\alpha$(J2000)$=5^{h} 17.6^{m}$, 
$\delta$(J2000)$=-69\arcdeg 2^{'}$ determined from the kinematics of the HI 
(Kim et al.\ 1998). The region within 1.7 kpc from the center (dashed line)
is completely covered by the current survey. The surface density is derived
by dividing the mass of the clouds by the area covered by the survey.
%The dashed line shows the best fitting linear equation of
%$\Sigma = 1.2 \times 10^{-3} d_{\rm cen} +3.9$.
(b) Distribution of the surface density 
along the position angle starting from the north to the east in counter clock-wise direction
with respect to the kinematic center as of (a).
}
\label{fig:sddist}
\end{figure}
\clearpage

\clearpage
\begin{figure*}
\epsscale{1.0}
\plotone{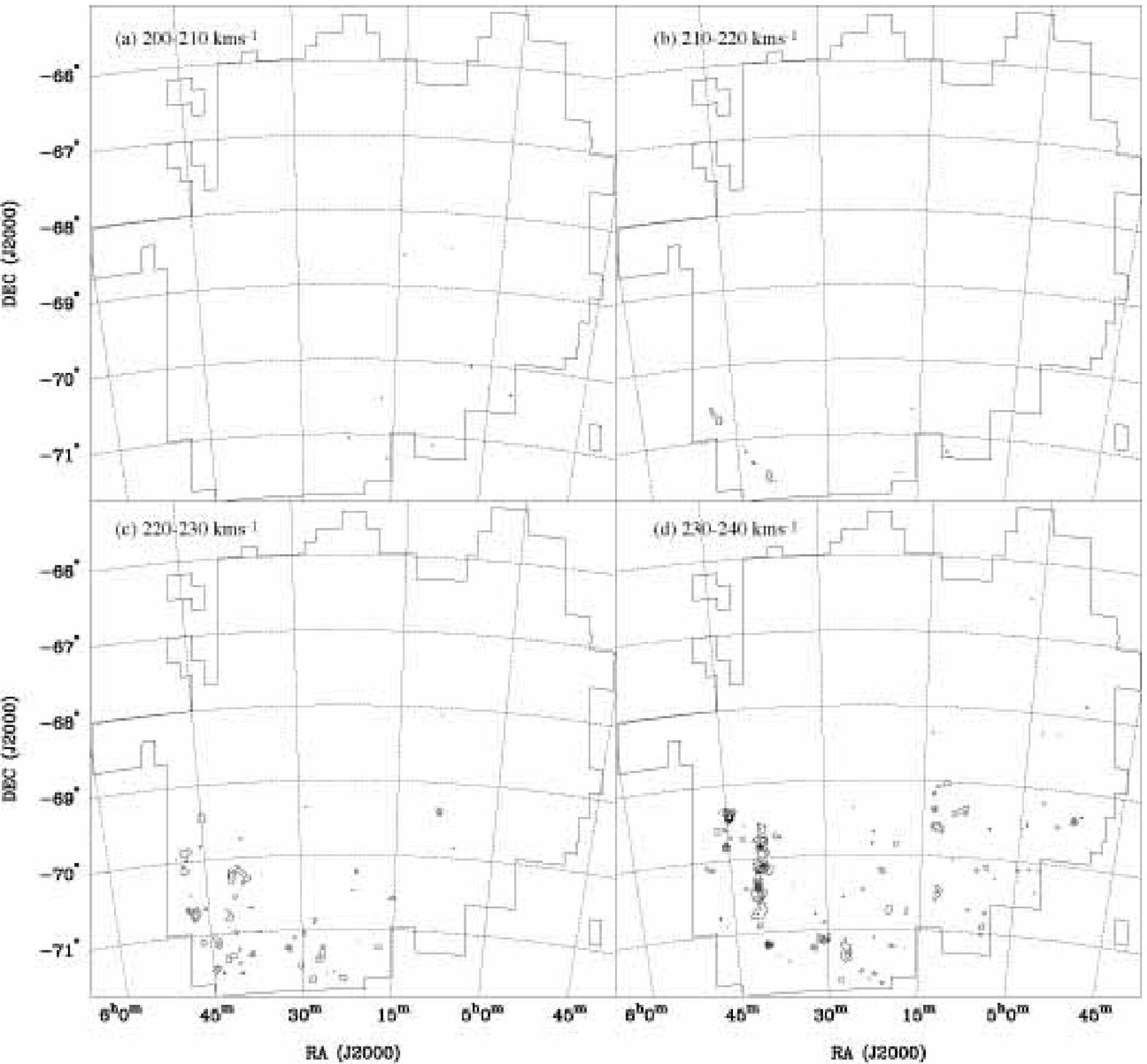}
\caption{Velocity-integrated maps of the $^{12}$CO emission integrated over
(a) $200 \le V_{\rm LSR} < 210$ km s$^{-1}$,
(b) $210 \le V_{\rm LSR} < 220$ km s$^{-1}$,
(c) $220 \le V_{\rm LSR} < 230$ km s$^{-1}$,
(d) $230 \le V_{\rm LSR} < 240$ km s$^{-1}$,
(e) $240 \le V_{\rm LSR} < 250$ km s$^{-1}$,
(f) $250 \le V_{\rm LSR} < 260$ km s$^{-1}$,
(g) $260 \le V_{\rm LSR} < 270$ km s$^{-1}$,
(h) $270 \le V_{\rm LSR} < 280$ km s$^{-1}$,
(i) $280 \le V_{\rm LSR} < 290$ km s$^{-1}$,
(j) $290 \le V_{\rm LSR} < 300$ km s$^{-1}$,
(k) $300 \le V_{\rm LSR} < 310$ km s$^{-1}$,
and (l) $310 \le V_{\rm LSR} < 320$ km s$^{-1}$, respectively. 
The contours are from 1 K km s$^{-1}$ with 
1 K km s$^{-1}$ intervals. The thin lines indicate the observed region.
}
\label{fig:ch1}
\end{figure*}
\clearpage

\begin{figure*}
\epsscale{1.0}
{\plotone{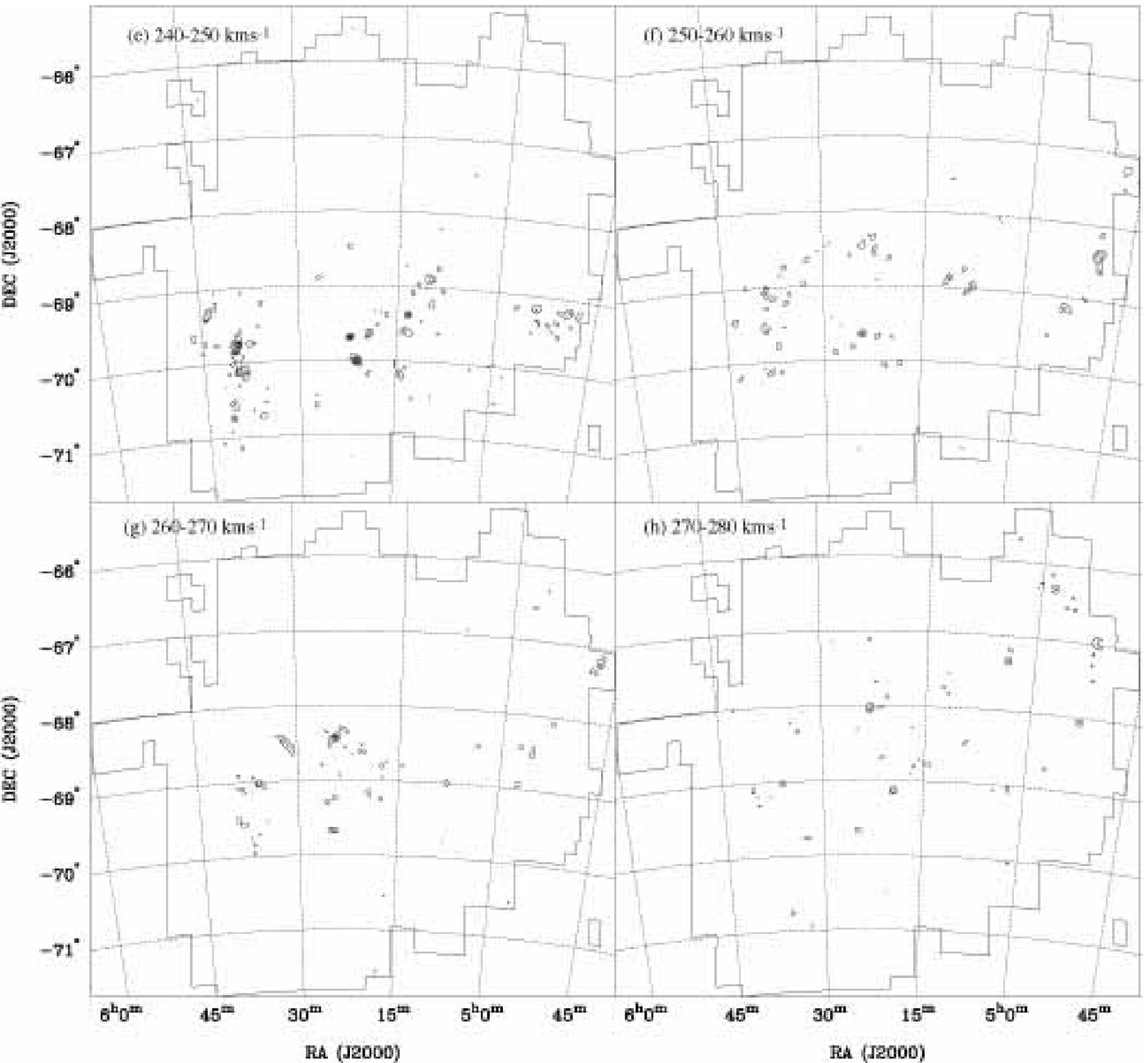}}\\[5mm]
\centerline{Fig. 5. --- {\it continued}}
\label{fig:ch2}
\end{figure*}
\clearpage

\begin{figure*}
\epsscale{1.0}
{\plotone{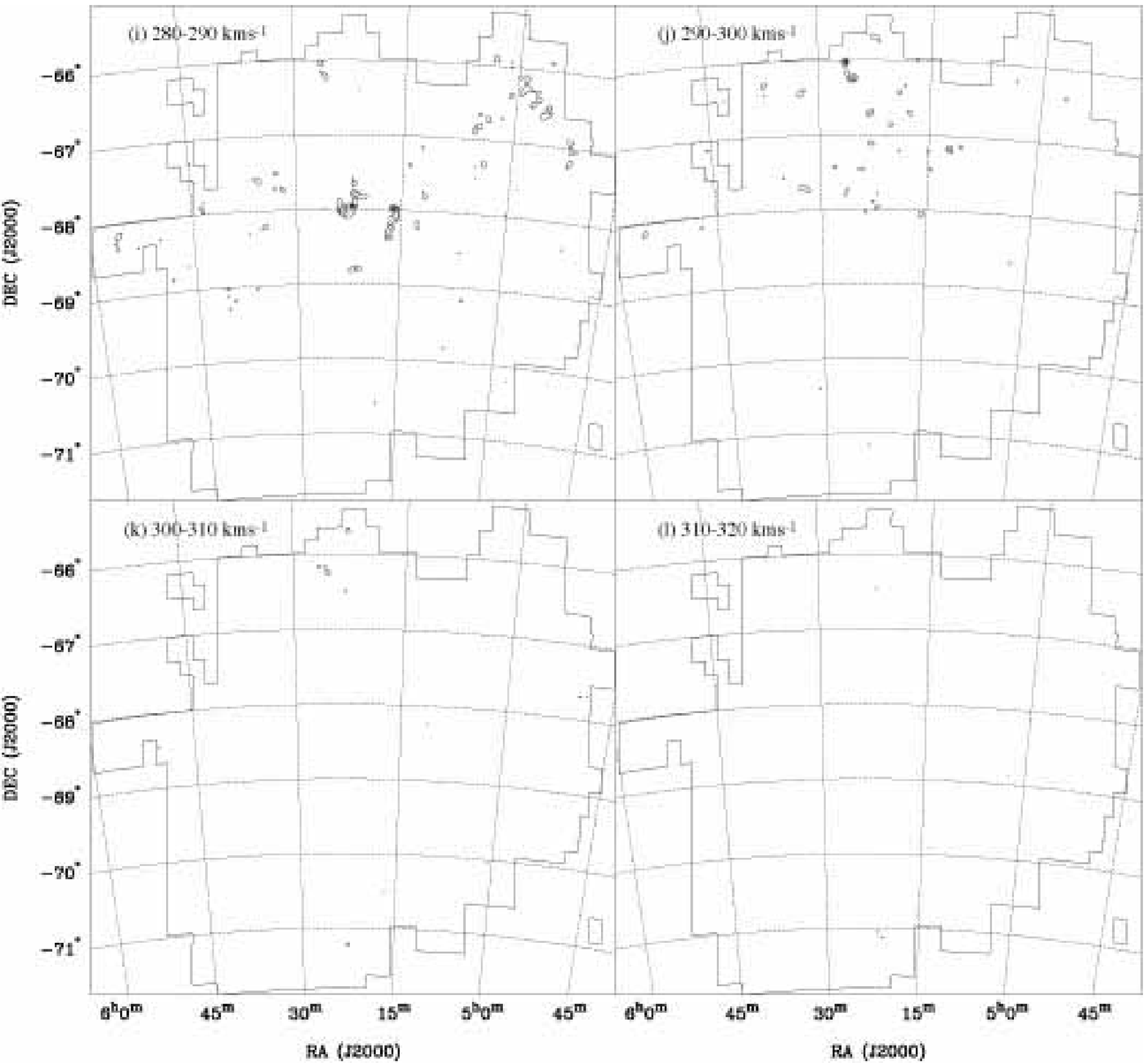}}\\[5mm]
\centerline{Fig. 5. --- {\it continued}}
\label{fig:ch3}
\end{figure*}
\clearpage

\begin{figure*}
\epsscale{1}
\plotone{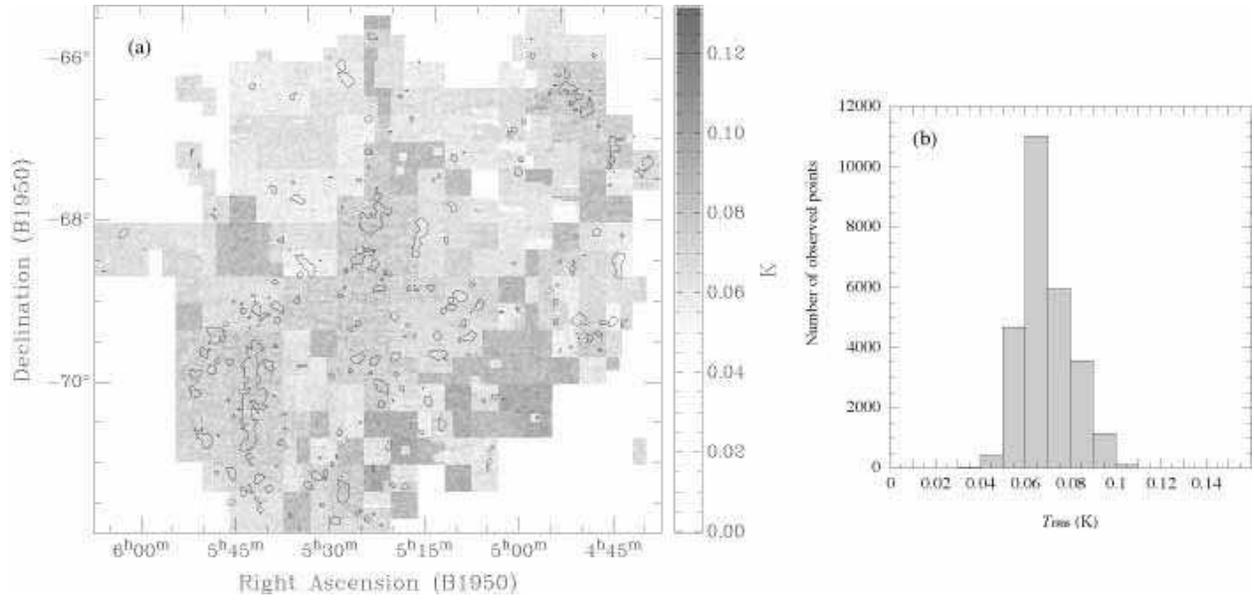}
\caption{
(a) Spatial distribution of the noise fluctuations per AOS channel of the current survey
and (b) frequency distribution of the noise fluctuations.
}
\label{fig:rms}
\end{figure*}
\clearpage

\clearpage
\begin{figure}
\epsscale{0.6}
\plotone{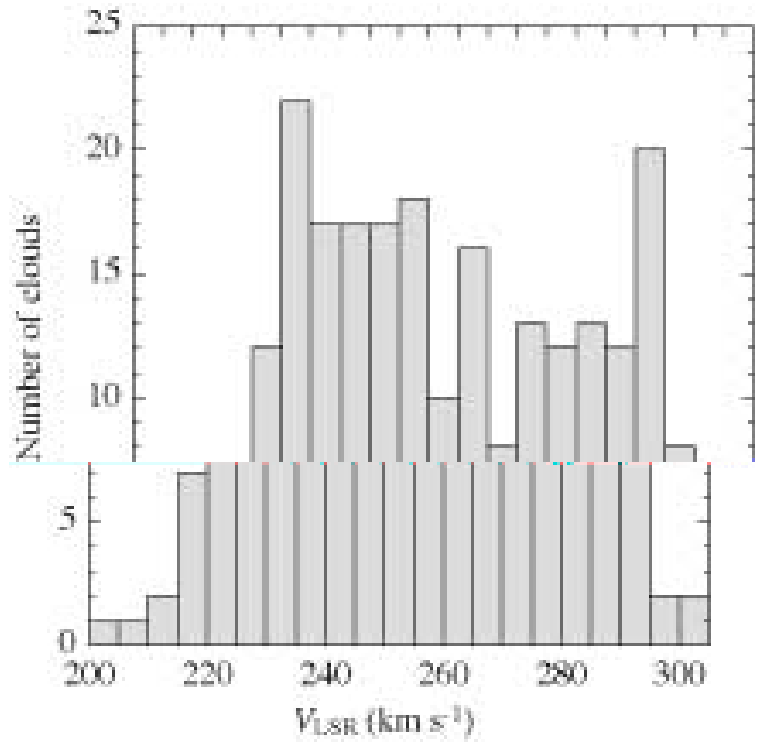}
\caption{
Frequency distribution of the $V_{\rm LSR}$ of the clouds.
}
\label{fig:vlsrhist}
\end{figure}

\clearpage
\begin{figure}
\epsscale{0.6}
\plotone{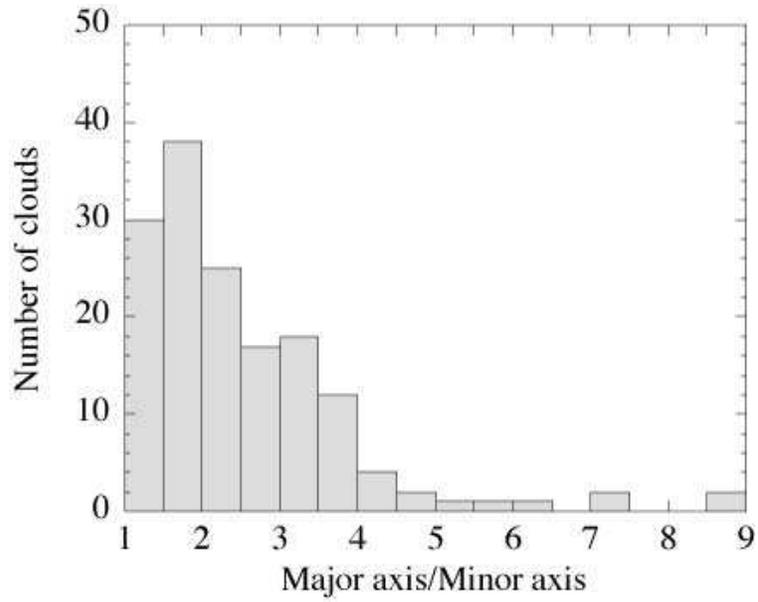}
\caption{
Distribution of the ratio of the major and minor axes of the clouds.
Both axes are derived by using the fitstoprops program
(Rosolowsky \& Leroy 2006) and then by de-convolving the NANTEN beam
(see section 3). 
}
\label{fig:manmin}
\end{figure}
%\clearpage

\clearpage
\begin{figure}
\epsscale{0.6}
\plotone{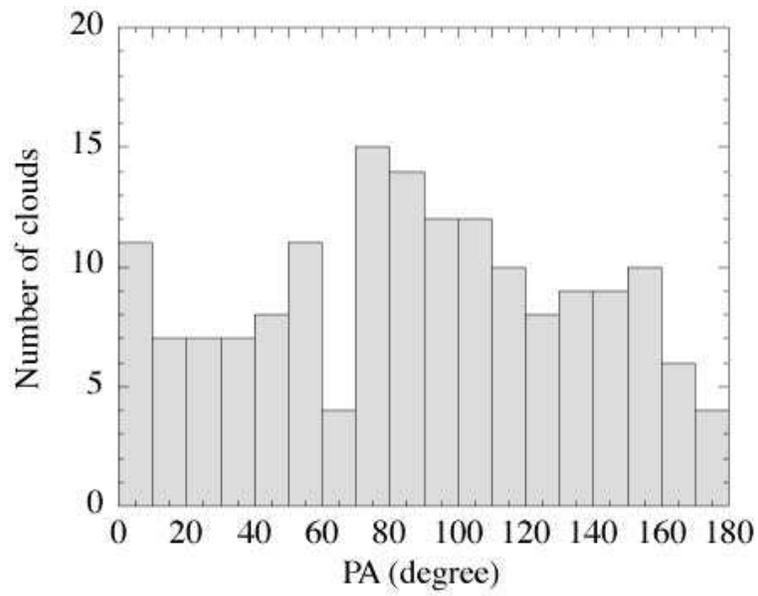}
\caption{
Position angle (PA) of the clouds.
PA is taken as $0\arcdeg$ toward the north
and then to the east in the counter-clockwise direction.
}
\label{fig:pa164}
\end{figure}

\clearpage
\begin{figure*}
\epsscale{1}
\plotone{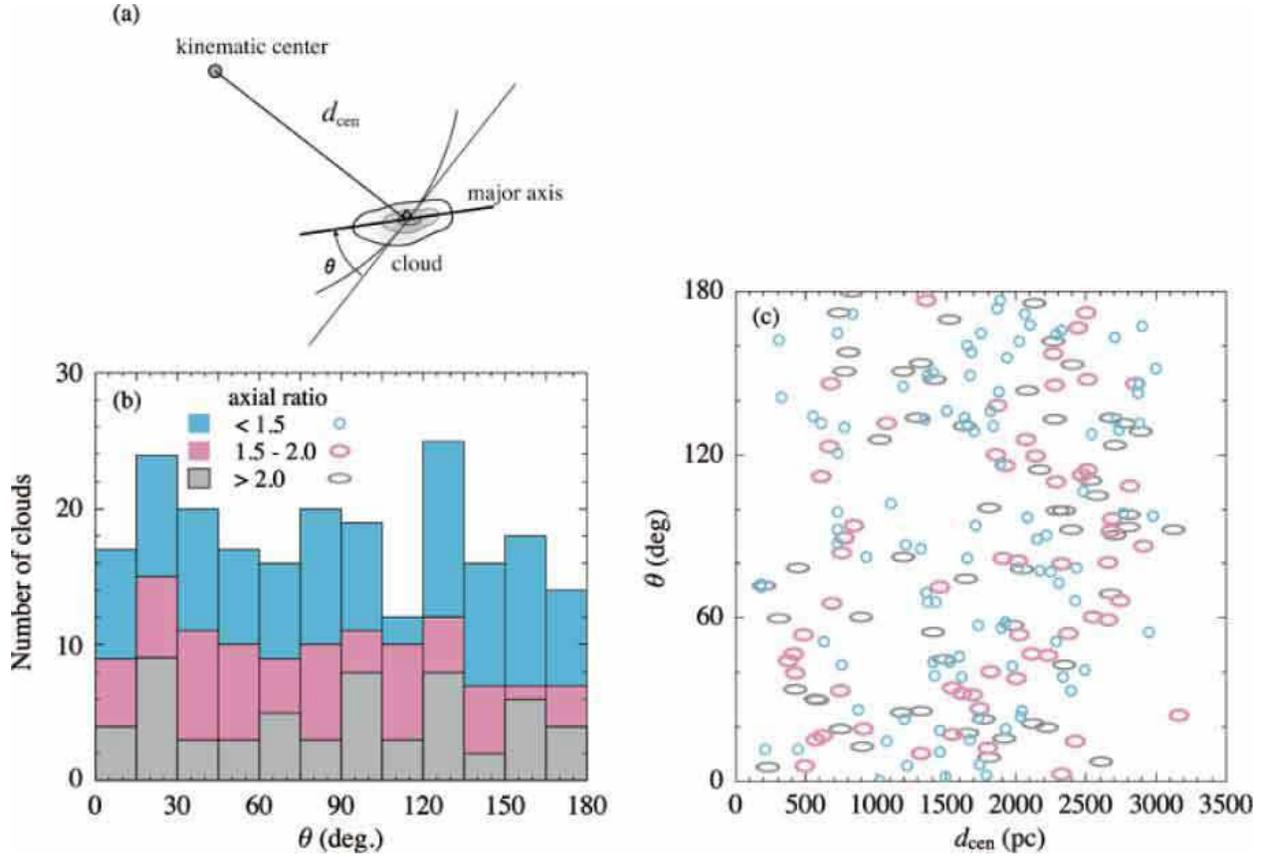}
\caption{
(a) Schematics to explain the angle, $\theta$, 
between the major axis of the GMC
and the tangent of the circle with a radius of the distance, $d_{\rm cen}$,
from the kinematic center derived from the HI by Kim et al.\ (1998).
(b) Distribution of $\theta$. 
The gray, pink, and blue show the number of the GMCs with the ratio of the major and minor
axes to be $> 2.0$, 1.5--2.0, and $< 1.5$, respectively
in electronic version.
(c)The angle, $\theta$, with respect to the distance of the GMCs
from the center derived from the HI kinematics by Kim et al.\ (1998).
The most elongated ellipse, ellipse, and circle show the GMCs with the ratio of the major and minor
axes to be $> 2.0$, 1.5--2.0, and $< 1.5$, respectively.
[The black ellipse, pink ellipse, and blue circle show
the GMCs with the ratio of the major and minor
axes to be $> 2.0$, 1.5--2.0, and $< 1.5$, respectively
in electronic version.]
}
\label{fig:theta}
\end{figure*}
%\clearpage

\clearpage
\begin{figure}
\epsscale{0.7}
\plotone{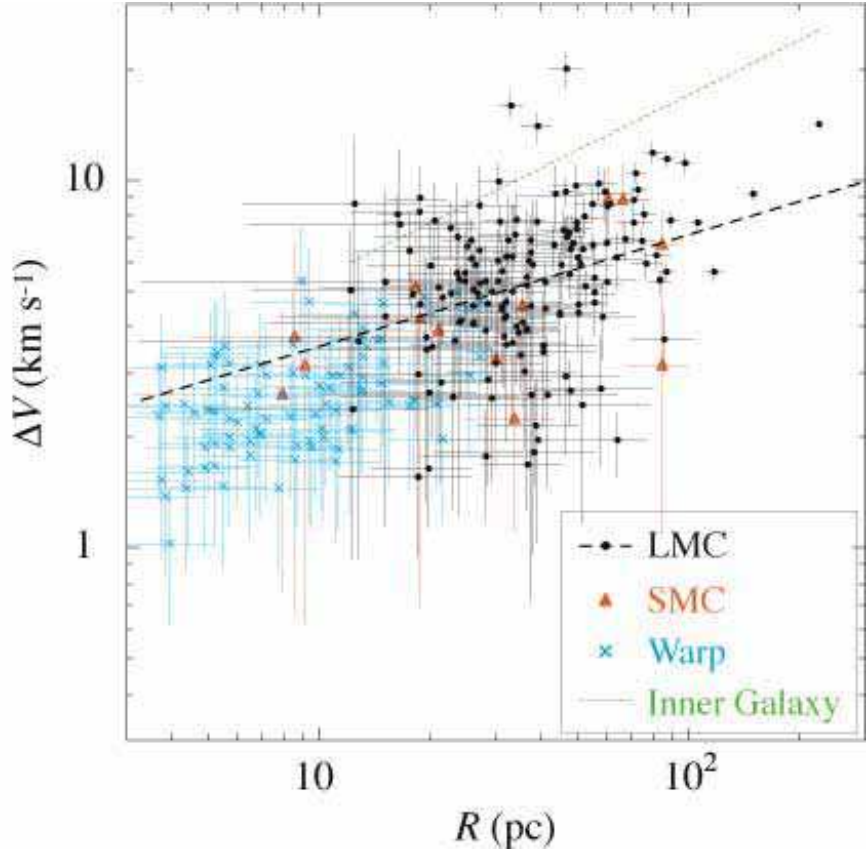}
\caption{
Line width-size relation for the GMCs in the LMC (filled circle).
Triangles [in red in electronic version] show the clouds in the SMC (Mizuno et al.\ 2001a)
and crosses [in blue in electronic version] in the Warp region (Nakagawa et al.\ 2005), respectively;
the clouds in these regions are re-identified 
by using the algorithm described in Rosolowskt \& Lorey (2006)
and the same parameters used to identify the clouds in the LMC.
The dashed line is the best fitting power law to the GMCs in the LMC,
$\Delta V = 1.3 R ^{0.2}$,
with Spearman rank coefficient of 0.3.
The dotted line [in green] is the relation found for the GMCs in the
inner Galaxy from Solomon et al.\ (1987) as an example.
}
\label{fig:rv}
\end{figure}
%\clearpage

\clearpage
\begin{figure}
\epsscale{0.7}
\plotone{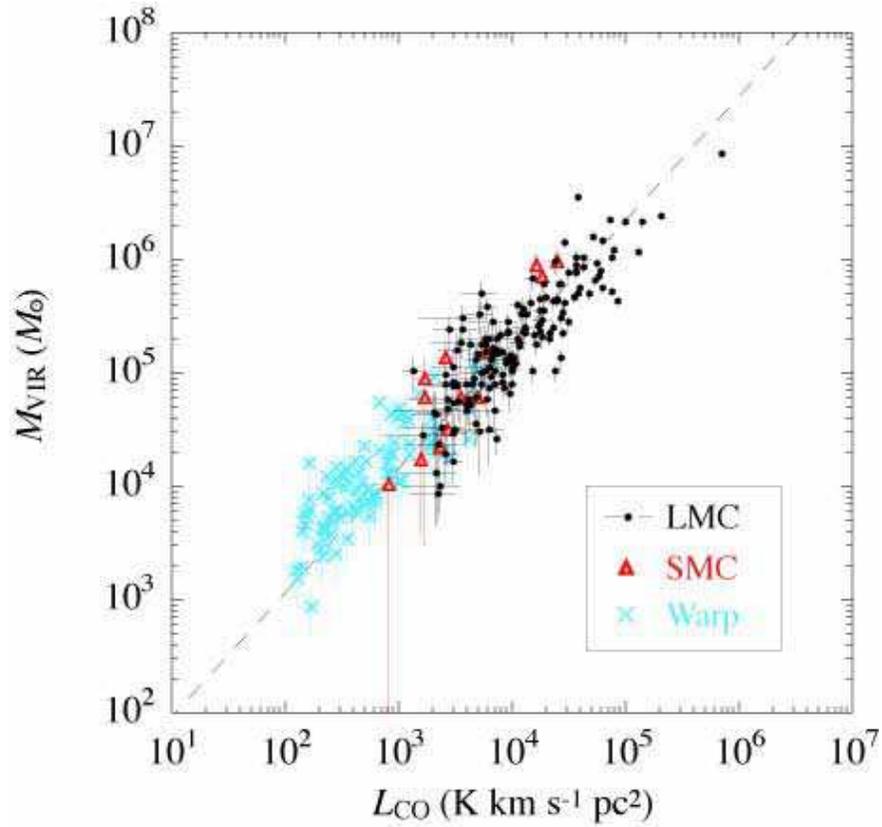}
\caption{
Plot of the virial mass, $M_{\rm VIR}$ of the GMCs as a function of luminosity, $L_{\rm CO}$
(filled circle). The line present a best fit to the data with slope of $1.1\pm 0.1$.
Triangles [in red in electronic version] show the clouds in the SMC (Mizuno et al.\ 2001a)
and crosses [in blue in electronic version] in the Warp region (Nakagawa et al.\ 2005), respectively;
the clouds in these regions are re-identified as for the LMC clouds by using the algorithm described by Rosolowsky\& Lorey (2006).}
\label{fig:mvl}
\end{figure}
%\clearpage

\clearpage
\begin{figure}
\epsscale{0.7}
\plotone{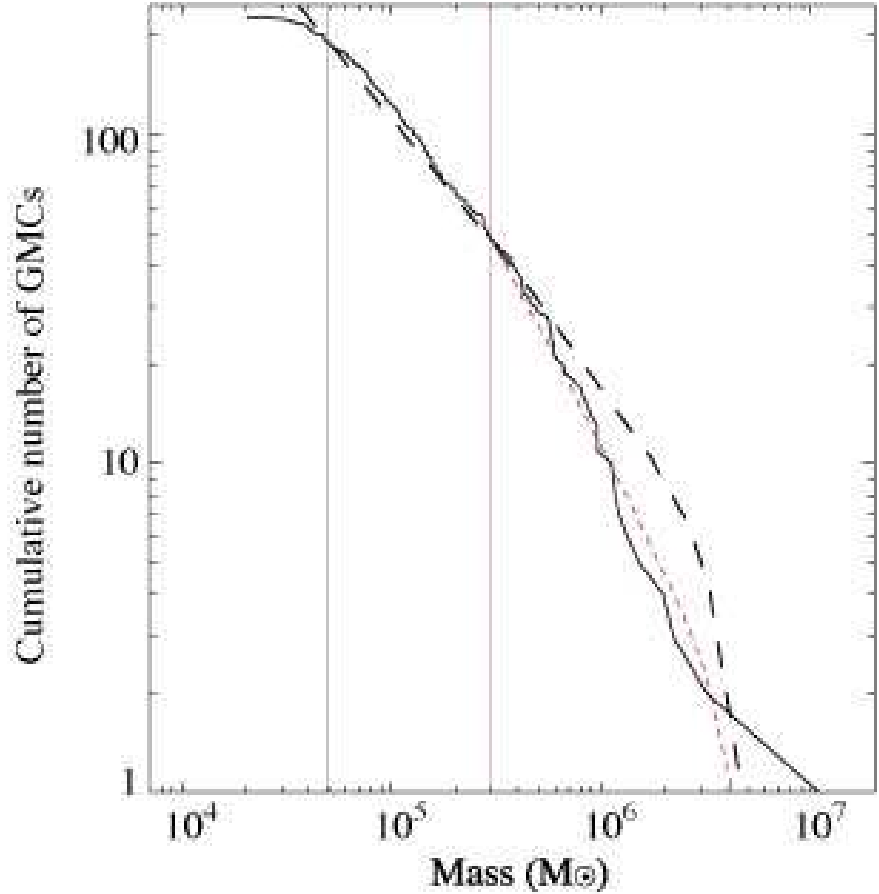}
\caption{
Cumulative mass spectra of the $M_{\rm CO}$ of the 230 GMCs.
The best-fitting power law to $M_{\rm CO}$ above the completeness limit 
$5 \times 10^{4} M_{\sun}$ (thin line)
is $N_{\rm cloud}$($\geq M_{\rm CO}$) = $6.6 \times 10^{5} M^{-0.75 \pm 0.06}-3.4$
and is indicated by a dashed line.
The [red] dotted line indicates the best-fitting power law, 
$N_{\rm cloud}$($\geq M_{\rm CO}$) = $1.3 \times 10^{8} M^{-1.2 \pm 0.2}-0.72$,
obtained by fitting $M_{\rm CO}$ of the 47 clouds with $M_{\rm CO} \geq 3 \times 10^{5}$
([red] thin line).
}
\label{fig:ms}
\end{figure}
%\clearpage

\clearpage
\begin{figure}
\epsscale{0.7}
\plotone{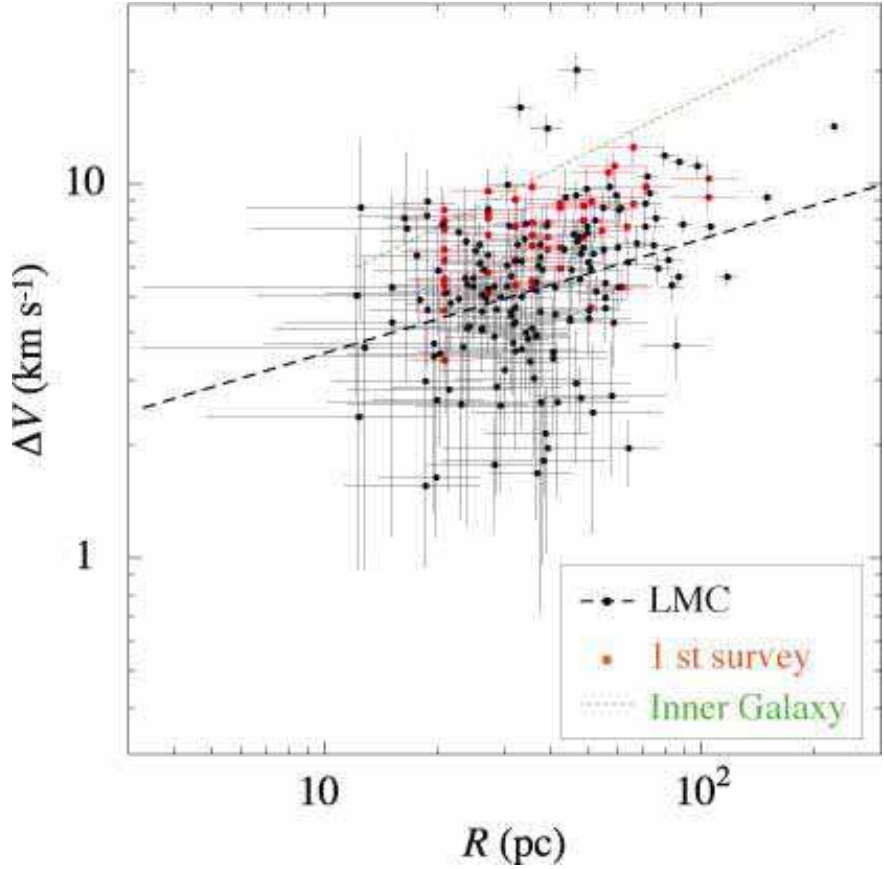}
\caption{
Line width-size relation of the Group A GMCs from the current survey (filled circle)
and 55 ``large clouds" from the 1st survey (red circle [in electronic version]).
The size of a cloud from the 1st survey was re-calculated 
from the catalog (Mizuno et al.\ 2001b) 
by subtracting the NANTEN beam size.
The dashed line indicates the best fitting power law of the GMCs from the 2nd LMC survey.
The dotted line is the relation found for the GMCs in the
inner Galaxy from Solomon et al.\ (1987) as an example.}
\label{fig:rdv1st}
\end{figure}

\clearpage
\begin{figure}
\epsscale{0.7}
\plotone{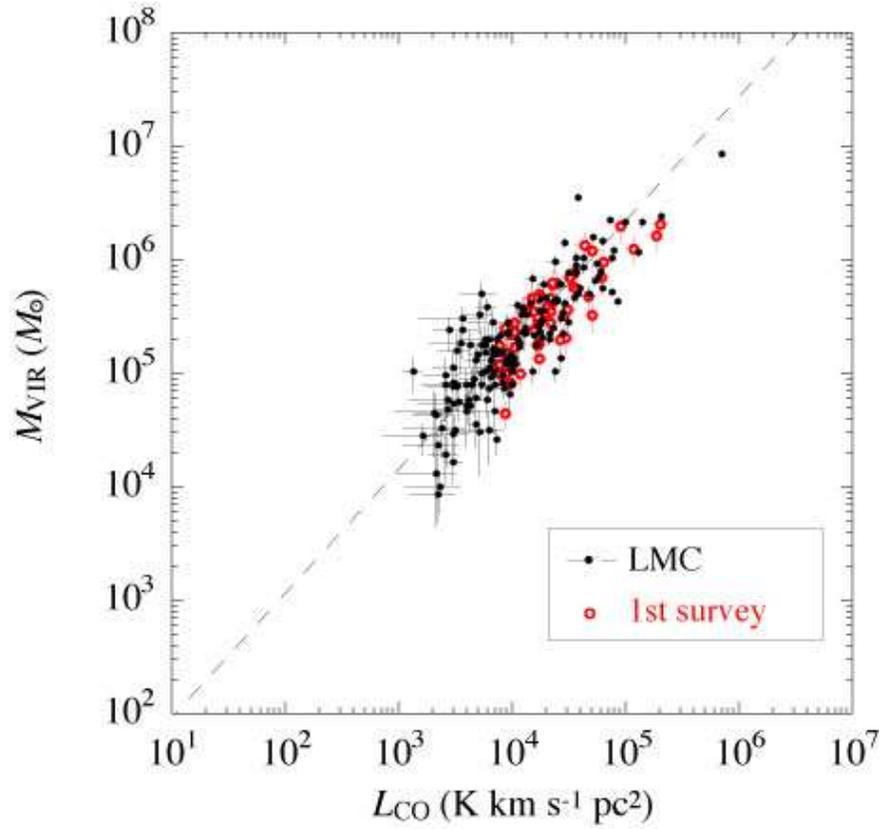}
\caption{
Plot of the virial mass, $M_{\rm VIR}$ of the 164 GMCs 
as a function of luminosity, $L_{\rm CO}$
from the current survey (filled circle). 
The line present a best fit to the data with slope of $1.1\pm 0.1$.
Red circles [in electronic version] are for the clouds from the 1st survey
(Fukui et al.\ 1999; Mizuno et al.\ 2001b).
The virial mass of a cloud from the 1st survey was re-calculated 
after subtracting the NANTEN beam size from the size of the cloud
shown in the catalog by Mizuno et al.\ (2001b).}
\label{fig:mvlco1st}
\end{figure}
%\clearpage

\clearpage
\begin{figure}
\epsscale{0.6}
\plotone{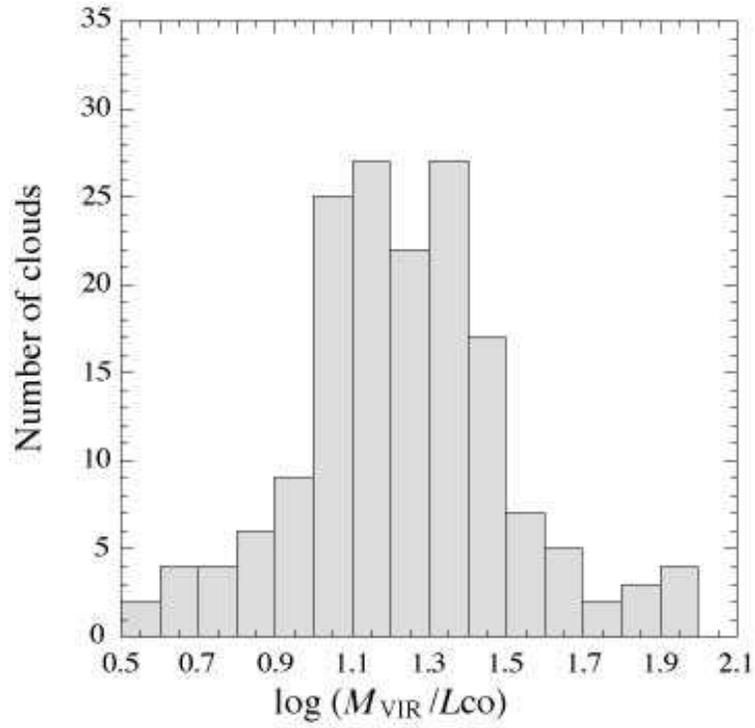}
\caption{
Histogram of log($M_{\rm VIR} / L_{\rm CO}$).
The mean of log($M_{\rm VIR} / L_{\rm CO}$) is $1.2 \pm 0.3$, corresponding to $X_{\rm CO} = (7 \pm 2)\times 10^{20}$ cm$^{-2}$(K km s$^{-1}$)$^{-1}$ }
\label{fig:mvlhist}
\end{figure}
%\clearpage

\clearpage
\begin{figure}
\epsscale{0.7}
\plotone{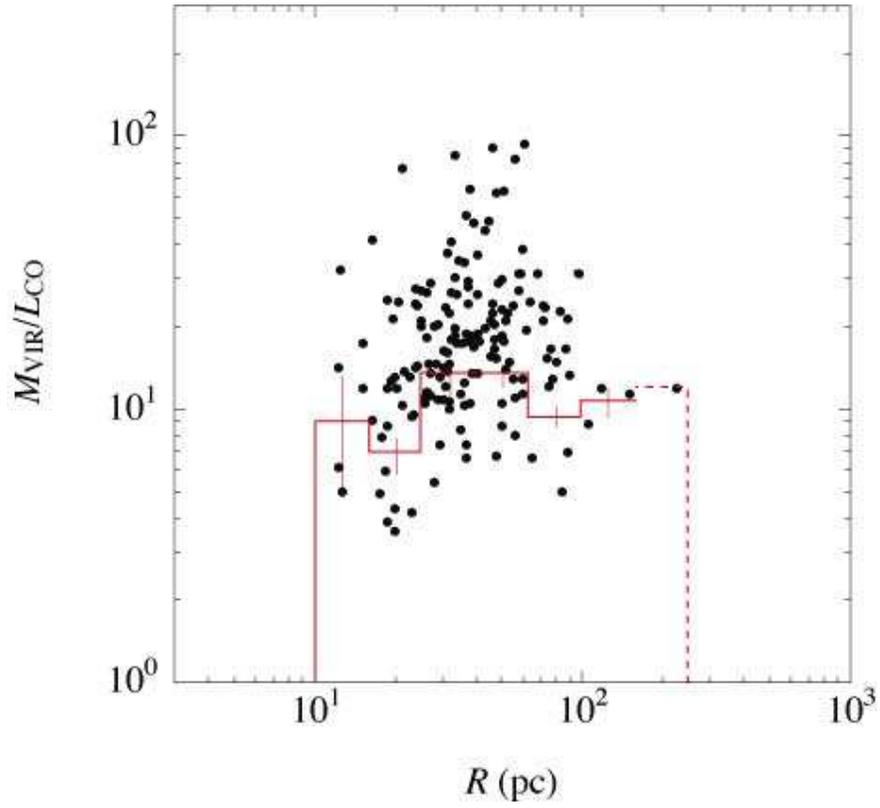}
\caption{Dependence of $M_{\rm VIR} / L_{\rm CO}$ on the size
of the Group A GMCs. 
The red line shows the weighted mean of the $M_{\rm VIR} / L_{\rm CO}$
in each bin of log ($R$)
; the last bin shown by the dashed-line includes only one point
of $M_{\rm VIR} / L_{\rm CO}$.
}
\label{fig:mvl_r}
\end{figure}

\clearpage
\begin{figure}
\epsscale{0.7}
\plotone{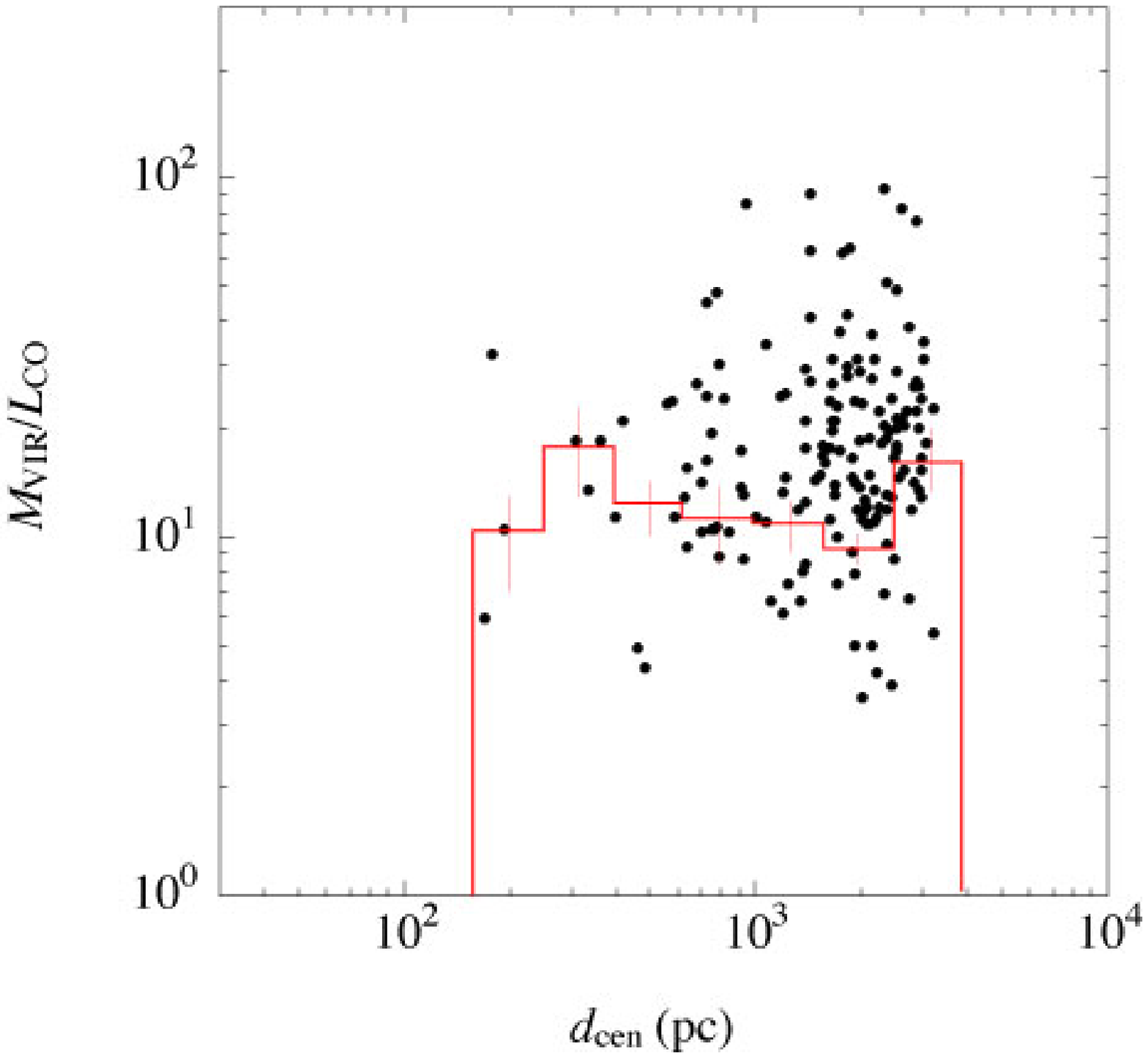}
\caption{
Dependence of $M_{\rm VIR} / L_{\rm CO}$ on the distance
of the GMCs from the center of the LMC, 
$\alpha$(J2000)$=5^{\rm h}17.6^{\rm m}$, 
$\delta$(J2000)$=-69\arcdeg2^{'}$ derived from the HI kinematics (Kim et al.\ 1998).
The red line shows the weighted mean of the $M_{\rm VIR} / L_{\rm CO}$
in each bin of log ($d_{\rm cen}$).
}
\label{fig:dcen_lco}
\end{figure}


\clearpage

\begin{thebibliography}{}
\bibitem[bm]{bm92}
Bertoldi, F.\ \& McKee, C.\ 1992, \apj, 395, 140

\bibitem[bc]{r3}
Bica, E., Claria, J. J., Dottori, H.,
Santos, J. F. C. Jr., \& Piatti, A. E.,
1996, \apjs, 102, 57


\bibitem[bt]{bt80}
Blitz, L. \& Thaddeus, P.\ 1980, \apj, 241, 676

\bibitem[ppv]{ppv}
Blitz, L., Fukui, Y., Kawamura, A.\, 
Leroy, A., Mizuno, N., \&
Rosolowsky, E. 2007, in Protostars and Planet V,
ed. B.\ Reipurth, D.\ Jewitt, \& K. Keil (University of Arizona) 951

\bibitem[Bloemen et al. (1996)]{bloe1996}
Bloemen,~J.~B.~G.~M., Strong,~A.~W.,  
Blitz,~L., Cohen,~R.~S., Dame,~T.~M., 
Grabelsky,~D.~A.,  
Hermsen,~W., Lebrun,~F., et al. 
1986, A\&A, 154, 25

\bibitem[Blum et al. 2006]{blum06}
Blum, R.\ D.\ et al.\ 2006, \aj, 132, 2034

\bibitem[ck1996]{ck96}
Caldwell, D.\ A.\ \& Kutner, M.\ L.\ 1996, \apj, 472, 611

\bibitem[ccg84]{ccg84}
Casoli, F.\, Combes, F.\, \& Gerin, M.\ 1984, \aap, 133,99

\bibitem[Cohen et al.\ 1988]{cohen88}
Cohen, R.\ S., Dame, T. M., Garay, G., Montani, J., Rubio, M., \& Thaddeus, P. 1988, \apj, 331, 95

\bibitem[Crawford, Jauncey, and Murdoch (1970)]{cra70}
Crawford, D. F., Jauncey, D. L., \& Murdoch, H. S.\ 1970, \apj, 162, 405

\bibitem[Dame et al.\ 1986]{dame86}
Dame, T. M., Elemegreen, R.\ S., Cohen, R.
S., \& Thaddeus, P. 1986, \apj, 305, 892

\bibitem[dem]{r8}Davies, R. D., Elliott, K. H., \&
Meaburn, J. 1976, MmRAS, 81, 89

\bibitem[Dickel et al. 2005]{dickel05}
Dickel, J.\ R., Mclntyre, V.\ J., Gruendl, R.\ A., \& Milne, D.\ K. 2005, \aj, 129, 790

\bibitem[Digel1996]{digel96}
Digel et al.\ 1996, \apj, 458, 561

\bibitem[Dobashi et al. (1996)]{dobashi1996}
Dobashi,~K., Bernard,~J.~P., \& Fukui,~Y. 1996, ApJ, 466, 282

\bibitem[dufour]{r10}
Dufour, R.\ J.\ 1984, in Structure and Evolution of the Magellanic Clouds,
ed.\ S.\ van den Bergh \& K.\ S.\ de Boer
(Dordrecht: Reidel Publishing Co.) 353


\bibitem[Filipovic 1996]{fil96}
Filipovic, M.\ D., Jones, P.\ A., White, G.\ L., \& Haynes, R.\ F.,
Meinert, D., Wielebincki, R., \& Klein, U.\ 1996, \aaps, 120, 77

\bibitem[Fukui 2007]{fukui2007}
Fukui, Y. 2007, 
in IAU Symp. 237, 
Triggered Star Formation in a Turbulent ISM,
ed. B. Elmegreen \& J. Palous (Cambridge)

\bibitem[Fukui 2008]{fukui2008}
Fukui, Y. et al. 2008,  in preparation (Paper III)

\bibitem[Fukui 2005]{fukui2005}
Fukui, Y. 2005, 
in IAU Symp. 227, 
Massive Star Birth: A Crossroads of Astrophysics,
ed. R. Cesaroni, M. Felli, E. Churchwell, \& M. Walmsley (Cambridge), 328

\bibitem[Fukui 2002]{fukui2002}Fukui, Y. 2002, in Modes of Star
Formation and the Origin of Field Populations, ASP Conference Proc. Vol. 285, 
ed. E.K.Grevel \& W. Brandner (San Francisco: ASP), 24

\bibitem[Fukui et al.\ 1999]{r14}
Fukui, Y., Mizuno, N., Yamaguchi, R.,
Mizuno, A., Onishi, T., Ogawa, H.,
Yonekura, Y., Kawamura, A., Tachihara,
K., Xiao, K., et al.\ 1999, \pasj, 51,
745 

\bibitem[Fukui et al.\ 2001]{Fukui01}
Fukui, Y., Mizuno, N., Yamaguchi, R.,
Mizuno, A., Onishi, T. 2001, \pasj, 53L,
41

\bibitem[Fukui Sakakibara 1992]{Fukui92}
Fukui, Y.\ \& Sakakibara, O.\ 1992, Mitsubishi electronic ADVANCE, 60, 11


\bibitem[Haynes et al.\ 1991]{haynes91}
Haynes et a.\ 1991, \aap, 252, 475

\bibitem[Henize (1956)]{heni1956}
Henize,~K.~G. 1956, ApJS, 2, 315

\bibitem[Heyer et al.\ (2001)]{hey01}
Heyer, M. H., Carpenter, J. M., \& Snell, R. L. 2001, \apj, 551, 852

\bibitem[h1988]{hodge88}
Hodge, P.\ 1988, \pasp, 100, 1051

\bibitem[hughes06]{hughes06}
Hughes et al.\ 2006, \apj, in press

\bibitem[i2]{israel86}
Israel, F.\ P., de Graauw, Th., van de Stadt, H.,
\& de Vries, C.\ P. 1986, \apj, 303, 186

\bibitem[i2]{israel03}
Israel, F.\ P., Johhansson, L.\ E.\ B, Rubio, M., Garay, G.,
de Graauw, Th., Booth, R.\ S., Boulanger, F., Kutner, M.\ L.,
Lequeux, J., \& Nyman, L.-A.\ 2003, \aap, 406, 817

\bibitem[ita]{ita04}
Ita, Y., et al.\ 2004, \mnras, 347, 720

\bibitem[j]{Johan94}
Johansson, L.\ E.\ B., Olofsson, H., Hjalmarson, A., Gredel, R., \& Black, J.\ H.\ 1994, \aap, 291, 89

\bibitem[j]{r23}
Johansson, L.\ E.\ B., Greve, A., Booth, R.\ S., Boulanger, F.,
Garay, G., Graauw, Th.\ de, Israel, F. P., Kutner, M.\ L.,
Lequeux, J., Murphy, D.\ C., Nyman, L.-A., \& Rubio, M. 1998, \aap, 331, 857

\bibitem[Kawamura et al. (1998)]{kawa1998}
Kawamura,~A., Onishi,~T., Yonekura,~Y., Dobashi,~K., Mizuno,~A.,
Ogawa,~H., \&
Fukui,~Y. 1998, ApJS, 117, 387

\bibitem[kh]{r24}
Kennicutt, R.\ C., Jr., \& Hodge, P.\ W.\ 1986, \apj, 306, 130
\bibitem[Kim98]{kim98}
Kim, S., Staveley-Smith, L., Dopita, M.\ A., Freeman, K.\ C., Sault, R.\ J.\, 
Kesteven, M.\ J.\ \&  McConnell, D.\ 1998, \apj, 503, 674

\bibitem[k]{r25}
Kim, S., Dopita, M.\ A., Staveley-Smith, L.,
\& Bessel, M.\ S. 1999, \aj, 118, 2797

\bibitem[Kim03]{kim03}
Kim, S., Staveley-Smith, L., Dopita, M.\ A., Sault, R.\ J., Freeman, K.\ C., 
Lee, Y., Chu, Y.-H.\ 2003, \apj, 148, 473

\bibitem[Koda et al.\ 2006]{koda06}
Koda, J., Sawada, T., Hasegawa, T., \& Scoville, N.\ Z.\ 2006,
\apj, 638, 191

\bibitem[koorn]{r26}
Koornneef, J.\ 1982, \aap, 107, 247

\bibitem[Kuter et al.\ 1997]{Kut97}
Kutner, M. L. et al.\ 1997 \aaps, 122, 255

\bibitem[Kuter \& Ulich]{KU81}
Kutner, M. \& Ulich, B.\ L.\ 1981, \apj, 250, 341

\bibitem[Larson 1981]{larson81}
Larson, R.\ B.\ 1981, \mnras, 194, 809

\bibitem[Lucke et al. (1970)]{luke1970}
Lucke,~P.~B., \&  Hodge,~P.~W. 1970, AJ, 75, 171
\bibitem[Luks et al. (1992)]{luks1992}
Luks,~Th., \& Rohlfs,~K. 1992, A\&A, 263, 41

\bibitem[McGee \& Milton 1966]{mm66}
McGee, R.\ X.\ \& Milton, J.\ A.\ 1966, AuPh, 19, 343


\bibitem[m]{r35}Meaburn, J. 1980, \mnras, 192, 365

\bibitem[Meixner et al.\ 2006]{m06}
Mexiner et al.\ 2006, \aap, 132, 2268

\bibitem[m1]{r36}
Mizuno, N., Rubio, M., Mizuno, A.,
Yamaguchi, R., Onishi, T., Fukui, Y.
2001a, \pasj, 53, L45

\bibitem[m2]{r37}Mizuno, N., Yamaguchi, R., Mizuno, A.,
Rubio, M., Abe, R., Saito, H., Onishi,
T., Yonekura, Y., Yamaguchi, N., Ogawa,
H., Fukui, Y. 2001b, \pasj, 53, 971

\bibitem[n]{r38}Nagahama, T., Mizuno, A., Ogawa, H.,
Fukui, Y. 1998, \aj, 116, 336

\bibitem[Nakagawa et al.\ 2005]{naka05}
Nakagawa, M., Onishi, T., Mizuno, A. \& Fukui, Y. 2005, \pasj, 57, 917

\bibitem[Ogawa et al.\ 1987]{ogawa87}
Ogawa, H., Mizuno, A., Hoko, H., Ishikawa, H., \& Fukui, Y. 1999, 
Int. J. Infrared and Millimeter Waves, 11, 717

\bibitem[o1993]{r43}
Ohta, K., Tomita, A., Saito, M., Sasaki, M., \& Nakai, N. 1993, \pasj, 45L, 21

\bibitem[Rolleston et al.\ 2002]{roll02}
Rolleston, W.\ R.\ J., Trundle, C.\ \& Dufton, P.\ L.\ 2002, \aap, 396, 53

\bibitem[Rosolowsky et al.\  2003]{ro03}
Rosolowsky, E., Engargiola, G., Plambeck, R., \& Blitz, L., 2003 \apj, 599, 258

\bibitem[Rosolowsky 2005]{ro05}
Rosolowsky, E.\ 2005 \pasp, 117, 1403

\bibitem[Rosolowsky \& Leroy 2006]{ro06}
Rosolowsky, E., \& Leroy, A.\ 2006 \pasp, 118, 590

\bibitem[Rubio et al.\ 1993]{rubio93}
Rubio, M., Lequeux, J., \& Boulanger, F.\ 1993 \aap, 271, 9

\bibitem[Rohlfs et al.\ 1984]{rohlfs84}
Rohlfs, K., Kreitschmann, J., \& Feitzinger, J.\ V.\ 1984, in IAU Symp. 108, 395

\bibitem[s]{s87}
Scoville, N.\ Z., Yun, M.\ S., Sanders, D.\ B., Clemens, D.\ P.\
\& Waller, W.\ H.\ 1987, \apjs, 63, 821

\bibitem[Snowden \& Petre, R. (1994)]{SP94}
Snowden, S.\ L.\ \& Petre, R.\ 1994, \apj, 436, 123

\bibitem[Solomon et al. (1987)]{sol87}
Solomon, P. M., Rivolo, A. R., Barrett, J., \& Yahil, A.\ 1987, \apj,
319, 730

\bibitem[Solomon \& Rivolo (1989)]{sol89}
Solomon, P. M.\ \& Rivolo, A.\ R.\ 1989, \apj,
339, 919

\bibitem[V\'{a}zquez-Semadeni, Ballesteros-Paredes, and Rodr\'{\i}guez
(1997)]{vaz97}
V\'{a}zquez-Semadeni, E., Ballesteros-Paredes, J., \& Rodr\'{\i}guez, L. F.\ 
1997, \apj, 474, 292

\bibitem[Wada, Spaans, and Kim (2000)]{wad00}
Wada, K., Spaans, M., \& Kim, S.\ 2000, \apj, 540, 797

\bibitem[Wada \& Norman (2001)]{wad01}
Wada, K.\ \& Norman, C.\ A.\ 2001, \apj, 547, 172

\bibitem[Wong \& Blitz 2002]{wong02}
Wong, T.\ \& Blitz 2002, \apj, 569, 157

\bibitem[ws]{ws90}
Wilson, C.\ D., \& Scoville, N. 1990, \apj, 363, 435

\bibitem[Yamaguchi et al.\ 2001a]{yama01a}
Yamaguchi, R., Mizuno, N., Onishi, T., Mizuno, A.,\&
 Fukui, Y. 2001a, \apj, 553, 185L

\bibitem[y1]{r51}Yamaguchi, R., Mizuno, N., Onishi, T.,
Mizuno, A., \& Fukui, Y. 2001b, \pasj, 53,
959
\bibitem[y2]{r52}Yamaguchi, R., Mizuno, N., Mizuno, A.,
Rubio, M., Abe, R.,Saito, H., Moriguchi,
Y., Matsunaga, L., Onishi, T., Yonekura, \&
Y., Fukui, Y. 2001c, \pasj, 53, 985

\bibitem[Yonekura et al.(1997)]{yone1997}
Yonekura,~Y., Dobashi,~K.,
Mizuno,~A., Ogawa,~H., \& Fukui,~Y. 1997, ApJS,
110, 21

\bibitem[Young et al.(1991)]{yong1991}
Young,~J.~S., \& Scoville,~N.~Z. 1991, 
ARA\&A, 29, 581  

\bibitem[Zaritsky et al.(2004)]{zaritsky2004}
Zaritsky, D., Harris, J., Thompson, I.\ B., \& Grebel, E.\ K.
2004, \aj, 128, 1606

\end{thebibliography}
\end{document}